\documentclass[draft]{agujournal2019}
\usepackage[utf8]{inputenc}
\usepackage{textgreek}
\usepackage{url} 
\usepackage{lineno}
\usepackage[inline]{trackchanges} 
\usepackage{soul}
\usepackage{amsmath}
\usepackage{float}
\usepackage{booktabs}
\usepackage{subcaption}
\usepackage{multirow}
\usepackage{bm}
\usepackage{rotating}
\usepackage{url}

\draftfalse
\usepackage{amssymb}
\usepackage{xcolor}

\journalname{Space Weather}

\begin{document}

%
%


\title{Reduced-Order Data Assimilation for Thermospheric Density Using Physics-informed SINDy$_c$ Models}

%
%




\authors{Sriram Narayanan\affil{1}\thanks{Current address: 1306 Evansdale Drive, Morgantown, WV 26506}, Daniele Sicoli\affil{1}, and Piyush Mehta\affil{1}}

\affiliation{1}{Department of Mechanical, Materials and Aerospace Engineering, West Virginia University, Morgantown, WV, USA.}





\correspondingauthor{Piyush Mehta}{piyushmukeshmehta@gmail.com}


\begin{keypoints}
\item A SINDy$_c$-AR reduced-order model propagates thermospheric density in a PCA latent space driven by solar and geomagnetic indices.
\item An Extended Kalman Filter assimilates in situ LEO density measurements, correcting the reduced-order latent state in real time.
\item Multi-satellite assimilation calibrates the latent state across altitudes and local times, improving density beyond single-track forecasts.
\end{keypoints}

%
%

%
%


\begin{abstract}
Orbit prediction and space situational awareness require accurate thermospheric mass density, which responds nonlinearly to solar and geomagnetic forcing across several orders of magnitude. Physics-based general circulation models resolve that response but are computationally expensive, while empirical models run cheaply and carry no time-evolving atmospheric state. An autoregressive Sparse Identification of Nonlinear Dynamics with control (SINDy$_c$-AR) model, derived from the Thermosphere-Ionosphere-Electrodynamics General Circulation Model (TIE-GCM), captures the dominant modes of variability and their dependence on the drivers at a fraction of the parent model's cost. An extended Kalman filter assimilates in situ observations from CHAMP, GRACE, GRACE-FO, GOCE, and Swarm into that model across several orbital configurations and geomagnetic conditions, with a linear DMDc model as reference. Assimilation reduces density error relative to open-loop prediction, most visibly during geomagnetic storms and under single-satellite coverage. The two reduced-order models perform comparably throughout. NRLMSIS~2.1 and HASDM can outperform the assimilated model far from the assimilated track, so results are framed as improvements over the open-loop forecast. An observation-calibrated global density record spanning August 2000 to December 2025 is released alongside this work.
\end{abstract}

\section{Plain Language Summary}
Predicting the density of the thermosphere, the upper layer of Earth's atmosphere, is important for tracking satellites and managing space operations. The thermosphere responds strongly and rapidly to changes in solar and geomagnetic activity, which makes accurate prediction hard. Physics-based models describe these processes in detail but are too slow for routine use, while simpler empirical models run quickly but track changing space-weather conditions less faithfully.
The method described here pairs a simplified, physics-informed atmospheric model with satellite measurements to produce observation-calibrated density estimates. The simplified model represents the main patterns of thermospheric behavior and how they change in response to space weather; a data assimilation technique updates the model using density observations from satellites such as CHAMP, GRACE, GRACE-FO, GOCE, and Swarm.
Incorporating satellite data reduces density errors relative to running the model open-loop, most clearly during disturbed conditions. The approach remains stable when observations are sparse and runs at a small fraction of the cost of full general-circulation propagation, supporting the update cadence used in operational drag prediction. A record of upper-atmosphere density from August 2000 through December 2025, produced with this method, is published alongside the paper.

\section{Introduction}

The thermosphere–ionosphere region of Earth's atmosphere, 100--980~km in the model grid used throughout this work (Section~\ref{sec:dataset}), sets the drag environment for most active satellites \cite{Prolss2004}. Thermospheric mass density drives the orbital evolution of low Earth orbit (LEO) objects, and so sets orbit prediction accuracy and space situational awareness \cite{Vallado2001}; since the early space age it has been inferred from orbital decay \cite{Sterne1958, KingHele1987, Picone2005}. Solar extreme ultraviolet (EUV) irradiance is the main driver, changing density near 400~km by an order of magnitude over a solar cycle, with annual, semiannual and diurnal cycles \cite{Paetzold1961, Jacchia1959} and geomagnetic storms adding more \cite{Jacchia1967, Liu2005a}. Small density errors compound into kilometer-scale orbital prediction errors over days, degrading conjunction assessment and reentry prediction for LEO spacecraft \cite{vallado2014critical}.

Empirical density models built from long-term satellite drag measurements, among them Jacchia–70 \cite{Jacchia1970b}, the Drag-Temperature Model (DTM) family \cite{Barlier1978b, Berger1998b, Bruinsma2003b}, NRLMSISE-00 \cite{Picone2002b} and Jacchia–Bowman 2006 \cite{Bowman2008a}, capture solar-cycle and seasonal variation and dominate operational orbit prediction; multi-satellite decay data and studies of 27-day solar rotational modulation have further characterized thermospheric variability \cite{Bowman2008b, Lean2006}. Accelerometers on missions such as CHAMP measure drag acceleration directly, resolving thermospheric structure and variability at LEO altitudes without relying on orbital decay over long arcs \cite{Liu2005a, Guo2007, Guo2008, Muller2009}.

Physics-based general circulation models give a more complete picture of thermospheric dynamics, especially under geomagnetic disturbance \cite{Sutton2018}. The Global Ionosphere–Thermosphere Model (GITM) \cite{Ridley2006}, TIE-GCM \cite{Qian2014} and the Whole Atmosphere Model (WAM) \cite{fuller2010whole} solve the coupled continuity, momentum and energy equations for neutral and ionized species, with state spaces that often exceed $10^5$ variables. Running them in real time is achievable: operational numerical weather prediction centers have done so with data assimilation for decades \cite{bauer2015quiet}, and the coupled Whole Atmosphere Model--Ionosphere Plasmasphere Electrodynamics system (WAM-IPE) has run operationally at NOAA's Space Weather Prediction Center since 2021 \cite{fang2022starlink}. The cost falls elsewhere: on the high-performance computing allocation each forecast cycle requires, which limits how many configurations can run, how fast a long historical interval can be reprocessed, and who can run the model at all.

Whether a physics-based model of this kind is operationally useful depends on what \emph{operational} is taken to mean, and the term covers two objectives that are often conflated. One is specification and forecast of the geospace environment, the objective WAM-IPE addresses, with part of its forecast lead time coming from solar wind measured at L1 and the rest from forecast $K_p$ and $F_{10.7}$~\cite{swpc_wamipe, ccmc_wamipe}; that capability needs the high-performance computing allocation described above, which neither a satellite operator's ground segment nor a spacecraft has, so it cannot itself support satellite operations. The other objective is a density model a satellite operator can run: at their own ground segment and, potentially, onboard the spacecraft for autonomous operations. Meeting it means giving up the physics-based model's full state space for something cheap enough to run there, which is what reduced-order modeling (ROM) provides.

ROM trades a bounded amount of physical fidelity for several orders of magnitude of compute reduction~\cite{Qian2014, Mehta2017}, compressing that output into a low-dimensional latent space, usually by proper orthogonal decomposition (POD) or principal component analysis (PCA)~\cite{Golub1970, Rowley2004, Jolliffe2011}. Empirical orthogonal function analyses of thermospheric density have characterized its dominant modes of variability~\cite{Matsuo2010} and built basis functions for calibrating semi-empirical models~\cite{Sutton2012}, the same compact-basis principle a dynamic ROM builds on~\cite{Mehta2017}; black-box compression with autoencoders is a growing alternative~\cite{rope_dim_red}.

Dimensionality reduction fixes the coordinates of the reduced state, but not the model that advances it in time; that model can be identified from trajectory data using operator-theoretic methods grounded in Koopman theory~\cite{koopman, Mezic2005}. Dynamic Mode Decomposition (DMD) computes a best-fit linear operator in the original state space, a finite-rank Koopman approximation on a linear observable subspace~\cite{Schmid2010, rowley2009}. With external inputs such as solar and geomagnetic forcing, DMD with control (DMDc) adds an explicit input term~\cite{Proctor2016}, giving $\mathbf{z}_{k+1} = \mathbf{A}\mathbf{z}_k + \mathbf{B}\mathbf{u}_k$, where $\mathbf{z}_k \in \mathbb{R}^r$ is the reduced state at step $k$, $\mathbf{u}_k$ the external drivers, and $\mathbf{A}$, $\mathbf{B}$ the state and input matrices identified from data, all defined in full in Section~\ref{sec:Math}.

The Sparse Identification of Nonlinear Dynamics (SINDy) framework~\cite{Brunton2016} takes a complementary approach: rather than a globally linear operator, it identifies a sparse set of coefficients $\boldsymbol{\Xi}$ so the dynamics are a linear combination of nonlinear basis functions drawn from a candidate library $\boldsymbol{\theta}$,
\begin{align}
\dot{\mathbf{z}} = \boldsymbol{\Xi}\,\boldsymbol{\theta}(\mathbf{z}, \mathbf{u}),
\end{align}
where $\boldsymbol{\theta}(\mathbf{z},\mathbf{u}) \in \mathbb{R}^{p}$ collects the $p$ candidate library functions evaluated at a single state and driver pair, and $\boldsymbol{\Xi} \in \mathbb{R}^{r \times p}$ holds the identified coefficients. The model is therefore linear in $\boldsymbol{\Xi}$, while nonlinearity enters through the choice of library functions in $\boldsymbol{\theta}$: the dynamics are not arbitrary nonlinear functions of the state, as in general nonlinear system identification, but a compact, interpretable expansion over a user-defined function space. The SINDy with control extension (SINDy$_c$)~\cite{Brunton2016b} adds driver-dependent and cross terms to capture state--driver interactions, and the autoregressive extension adopted here, SINDy$_c$-AR, adds lagged state and driver values, so that temporal lags act as lifting coordinates in the Koopman sense~\cite{arbabi2017, pan2020}. For completeness, neural networks are also used to model the dynamics of the reduced state~\cite{rope_dim_red_2}; the same computational economy that makes ROM attractive also makes an ensemble of dynamics models tractable at an emulator's compute budget rather than a general circulation model's, with the spread across those models providing uncertainty quantification and evolving ROM into a reduced order probabilistic emulator (ROPE)~\cite{ml_rope, thermosphere_problem, rope_ensemble}. This work uses both: SINDy$_c$-AR as the nonlinear reduced-order model, and DMDc as the linear reference it is compared against.

A reduced-order model identified this way still runs open-loop: it propagates from its initial condition and its drivers, with nothing to pull it back when it drifts from the atmosphere it represents. Data assimilation supplies that correction, combining model predictions with observations to reduce state uncertainty \cite{Shim2014}. In the thermosphere it has been applied to neutral density and satellite drag using Kalman filtering and ensemble-based methods \cite{Codrescu2004, Codrescu2018, FullerRowell2004, Matsuo2012, Matsuo2013a, Matsuo2013b, Sutton2018}, and \citeA{pedatella2020assimilation} assimilate ionospheric and thermospheric observations in a whole-atmosphere framework (WACCMX+DART), where adjusting electron density and neutral composition together improves short-term forecasts. But assimilation in high-dimensional models is computationally demanding. Ensemble methods keep it feasible without ever forming the full forecast covariance: they carry it as a low-rank square root built from the ensemble members and update it in that form \cite{tippett2003ensemble}, and covariances for selected variables or regions can be computed from the members when they are needed. The cost is that the covariance is estimated from a finite sample, so it carries sampling error and its rank is limited by the ensemble size. Coupling reduced-order models with data assimilation \cite{Mehta2018a, Mehta2018b} gives efficient state estimation without adjusting the external drivers directly, though it does not resolve how well model error is characterized: process noise still needs tuning, and the reduced model inherits whatever biases its parent carries. What it changes is the size of the problem: a covariance of order $10^1$--$10^2$ states can be stored and propagated explicitly, with no sampling error. The propagation is exact only for the reduced, locally linearized model, not for the full nonlinear uncertainty distribution.

This work builds on two lines of prior effort. \citeA{Mehta2018a} and \citeA{Mehta2018b} coupled a linear (DMDc) ROM of TIE-GCM with a Kalman filter to assimilate accelerometer-derived densities, and \citeA{sindy0} introduced a SINDy$_c$-based nonlinear ROM emulator of TIE-GCM without a data-assimilation layer, which \citeA{sicoli2026sindyar} extended to an autoregressive formulation on a grid reaching 980~km. This work couples the two: a quasi-physical ROM built with both DMDc and an autoregressive SINDy$_c$ formulation, trained on TIE-GCM simulations, with an Extended Kalman Filter (EKF) assimilating in situ mass density from CHAMP, GRACE, GRACE-FO, GOCE and Swarm. It is evaluated on withheld orbits across geomagnetic conditions and single-, dual- and multi-satellite configurations, against open-loop reduced-order predictions, NRLMSIS~2.1~\cite{Emmert2021, Emmert2022} and the High Accuracy Satellite Drag Model (HASDM)~\cite{storz2005high} where its 2000--2025 coverage permits.

This paper contributes: (i) an EKF that operates on the companion-form lifting of the SINDy$_c$-AR emulator, so the filter propagates the cross-lag covariances the autoregressive structure implies; (ii) an analytical Jacobian propagation for the nonlinear library, keeping the error covariance consistent with the first-order expansion of the AR process so the process noise need not absorb linearization error; (iii) a log-density observation model whose measurement Jacobian is state-independent, giving a well-conditioned update across the full dynamic range of density; (iv) a multi-satellite extension that assimilates geometrically distinct tracks simultaneously in the shared latent space; and (v) an assimilated thermospheric density dataset spanning 2000--2025, released with this work.

The reduced state is what makes this affordable: a 60-state EKF update costs on the order of a $60 \times 60$ covariance propagation rather than a computing allocation, so the covariance that carries the uncertainty on the density is propagated explicitly through the Jacobian of the reduced-order model rather than estimated from an ensemble (Section~\ref{subsubsec:cov_tuning}). The identified model integrates roughly $10{,}000$ times faster than the TIE-GCM run it was trained on. That serves the satellite operator's objective in three ways: real-time operation as a density component inside a drag or orbit-determination pipeline, onboard or at the ground segment; research use without a computing allocation, since an experiment on the 60-state filter runs on a single workstation; and reconstruction, since a multi-year assimilation pass costs a fraction of the equivalent general circulation model run and can be reprocessed as measurements or drivers are revised. Many satellite operators already mitigate density-model bias by estimating a ballistic coefficient or drag scale factor during orbit determination~\cite{vallado2026atmos}; this framework corrects a state shared across satellites rather than a per-satellite scalar, complementing that practice rather than replacing it.

The dataset released with this work, described fully in Section~\ref{sec:long_term_dataset}, is a continuous observation-calibrated global density record covering August 2000 to December 2025, built by running the filter forward across the period and ingesting whatever subset of CHAMP, GRACE, GRACE-FO, GOCE and Swarm measurements is available at each epoch. It ships in HDF5 format as the assimilated and open-loop reduced-order states on a 60-second time base, with a per-step log of contributing satellites, filter-consistency diagnostics and a notebook that reconstructs the density field on the full grid.

The remainder of this paper is organized as follows. Section~\ref{sec:Math} formulates the reduced-order models and the data assimilation framework. Section~\ref{sec:methodology} covers the data sources, preprocessing and filter tuning. Section~\ref{sec:results} presents results across the assimilation scenarios. Section~\ref{sec:long_term_dataset} describes the released dataset, and Section~\ref{sec:conclusion} summarizes the contributions and outlines future work.

\section{Mathematical Background}
\label{sec:Math}

The estimation problem addressed below is set by the dimension of the state: a global TIE-GCM \cite{Qian2014} snapshot on the standard latitude--longitude--altitude grid contains $72 \times 36 \times 45 = 116{,}640$ spatial degrees of freedom, which is what the reduction and the filter are built around.

The framework developed below combines three components. PCA compresses the high-dimensional thermospheric state into a low-dimensional latent representation that retains the dominant modes of global density variability. The latent state $\mathbf{z}_k$ and all associated dynamics operate in this linear reduced-order space throughout. The SINDy$_c$-AR model operates within this latent space to identify a compact representation of the latent-state evolution under external solar and geomagnetic forcing. An EKF operates in the same latent space to assimilate onboard satellite density measurements, correcting forecast errors and propagating uncertainty estimates alongside the state. The full thermospheric density field is reconstructed from the filtered latent state at any time via the inverse PCA projection.

The rest of this section builds on prior work~\cite{sindy0, sicoli2026sindyar}. Section~\ref{sec:dataset} defines the thermospheric state representation and describes the TIE-GCM training and testing datasets. Section~\ref{sec:emulation} details the PCA-based dimensionality reduction. Sections~\ref{subsec:sindy_library} and~\ref{subsec:sindy_ar} develop the SINDy$_c$ library construction and the SINDy$_c$-AR discrete-time dynamical model, respectively. Section~\ref{subsec:ekf} formulates the EKF for single-satellite assimilation, while Section~\ref{subsec:multi_sat_ekf} extends this to simultaneous multi-satellite assimilation.

\subsection{Thermospheric State and Training Dataset}
\label{sec:dataset}

Thermospheric density outputs from the TIE-GCM \cite{Qian2014} provide the training data, following previous data-driven work \cite{Mehta2018b, ml_rope, rope_dim_red}. The thermospheric state is discretized as a three-dimensional array
\begin{align}
\mathbf{x}_{\text{full}} \in \mathbb{R}^{n_\text{LT} \times n_\text{LAT} \times n_\text{ALT}},
\end{align}
where $n_\text{LT}=72$ denotes the number of local time bins, $n_\text{LAT}=36$ the number of latitude bins, and $n_\text{ALT}=45$ the number of altitude levels spanning 100--980~km. These indices correspond to the ranges
\begin{align}
\text{LT} \in [0,\,24)\ \text{hours}, \quad
\text{LAT} \in [-87.5^\circ,\,87.5^\circ], \quad
\text{ALT} \in [100,\,980]\ \text{km}.
\end{align}
The thermosphere is primarily forced by two external drivers: the solar 10.7~cm flux $F_{10.7}$ (in solar flux units, sfu), a proxy for solar extreme-ultraviolet radiation, and the planetary geomagnetic index $K_p$, which characterizes the level of geomagnetic activity. Following standard conventions, geomagnetic conditions are classified as quiet ($K_p < 3$), moderate ($3 \le K_p < 5$), strong ($5 \le K_p < 7$), or storm ($K_p \ge 7$).

To represent the periodic nature of solar and seasonal forcing, universal time (UT) and day of year (DOY) are encoded as circular features,
\begin{align}
t_1 &= \cos\!\left(\frac{2\pi\,\text{UT}}{24}\right), &
t_2 &= \sin\!\left(\frac{2\pi\,\text{UT}}{24}\right), \\
t_3 &= \cos\!\left(\frac{2\pi\,\text{DOY}}{365.25}\right), &
t_4 &= \sin\!\left(\frac{2\pi\,\text{DOY}}{365.25}\right).
\end{align}
This sine--cosine encoding ensures continuity across the boundaries of each 24-hour day and each calendar year, avoiding artificial discontinuities in the input feature space. The SINDy$_c$-AR emulator follows \citeA{sicoli2026sindyar}: it is trained on TIE-GCM output from 2002--2009 and tested on 1996--2001, because the 2002--2009 interval covers a wider range of solar and geomagnetic conditions. \citeA{sindy0} trained their SINDy$_c$ model the other way round, on 1996--2001, and tested it on 2002--2008.

\subsection{Dimensionality Reduction via PCA}
\label{sec:emulation}

The full thermospheric state snapshot at time step $k$, written as a column vector $\mathbf{x}_{\text{full},k}$, resides in a high-dimensional space of dimension $d = n_\text{LT}\,n_\text{LAT}\,n_\text{ALT}$. Collecting the $m$ hourly training snapshots as columns forms the state matrix $\mathbf{X}_{\text{full}} = [\mathbf{x}_{\text{full},1},\ldots,\mathbf{x}_{\text{full},m}] \in \mathbb{R}^{d \times m}$, where $m$ is the number of snapshots in the 1996--2009 training record and $k \in \{1,\ldots,m\}$ indexes time. PCA projects the full state onto an $r$-dimensional latent subspace. Each snapshot is first mean-centred by subtracting the mean state $\boldsymbol{\mu}_0 \in \mathbb{R}^d$, defined as the time average of the training snapshots, $\boldsymbol{\mu}_0 = \tfrac{1}{m}\sum_{k=1}^{m} \mathbf{x}_{\text{full},k}$. This is a full-state quantity in the original density units, not a latent-space quantity: it is the mean thermospheric state over the training period, that is the time-averaged density at every grid point. Mean-centring gives
\begin{align}
\mathbf{z}_k = \mathbf{W}^\top (\mathbf{x}_{\text{full},k} - \boldsymbol{\mu}_0), \quad \mathbf{z}_k \in \mathbb{R}^r, \quad r \ll d,
\end{align}
where $\mathbf{W} \in \mathbb{R}^{d \times r}$ contains the leading $r$ principal components and $\mathbf{z}_k$ represents deviations from the mean thermospheric state in the latent basis. In this work, $r = 10$ principal components are retained, enough to represent the large-scale spatial structure of the thermospheric density field while reducing the effective state dimension by several orders of magnitude. The choice of $r$ follows \citeA{sindy0} and was made by trial and error, balancing compact representation against variance retention. The full-state estimate is recovered via the inverse projection
\begin{align}
\hat{\mathbf{x}}_{\text{full},k} = \mathbf{W}\,\hat{\mathbf{z}}_k + \boldsymbol{\mu}_0.
\end{align}

\subsection{Library Construction and SINDy$_c$-AR}
\label{subsec:sindy_library}

With the thermospheric state represented in latent coordinates, the dynamics are modeled as
\begin{align}
\dot{\mathbf{z}}(t) = f\!\left(\mathbf{z}(t),\,\mathbf{u}(t)\right),
\end{align}
where $\mathbf{z}(t) \in \mathbb{R}^r$ is the latent state and $\mathbf{u}(t) \in \mathbb{R}^{n_\text{drivers}}$ is the vector of external drivers. The SINDy$_c$ framework \cite{Brunton2016, Brunton2016b} assumes that $f$ can be expressed as a sparse linear combination of candidate basis functions drawn from a user-defined library. As established in the original SINDy formulation~\cite{Brunton2016}, the resulting model is linear in the identified coefficients $\boldsymbol{\Xi}$, while nonlinearity enters through the choice of library functions. This is distinct from general nonlinear system identification: the dynamics are expressed as $\dot{\mathbf{z}} = \boldsymbol{\Xi}\,\boldsymbol{\theta}(\mathbf{z},\mathbf{u})$, where $\boldsymbol{\theta}$ is a fixed nonlinear feature map and $\boldsymbol{\Xi}$ is the coefficient matrix solved for by regression.

Throughout this section, a lowercase bold symbol denotes a quantity at a single time step ($\mathbf{z}_k \in \mathbb{R}^r$, $\mathbf{u}_k \in \mathbb{R}^{n_\text{drivers}}$, $\boldsymbol{\theta}(\mathbf{z}_k,\mathbf{u}_k) \in \mathbb{R}^{p}$), and the corresponding uppercase bold symbol denotes the matrix whose columns are those quantities over all $m$ training time steps ($\mathbf{Z} \in \mathbb{R}^{r \times m}$, $\mathbf{U} \in \mathbb{R}^{n_\text{drivers} \times m}$, $\boldsymbol{\Theta}(\mathbf{Z},\mathbf{U}) \in \mathbb{R}^{p \times m}$). With this convention the single-snapshot and matrix forms of the dynamics carry the coefficient matrix on the same side, $\dot{\mathbf{z}} = \boldsymbol{\Xi}\,\boldsymbol{\theta}(\mathbf{z},\mathbf{u})$ and $\dot{\mathbf{Z}} = \boldsymbol{\Xi}\,\boldsymbol{\Theta}(\mathbf{Z},\mathbf{U})$, so the ordering of the matrix product is fixed by the shapes and no transposition is implied between them.

The library matrix is constructed by evaluating a set of candidate functions over the available data:
\begin{align}
\boldsymbol{\Theta}(\mathbf{Z}, \mathbf{U}) =
\begin{bmatrix}
1 \\
z_i \\
z_i z_j \\
u_a \\
u_a u_b \\
z_i u_a \\
\vdots
\end{bmatrix}, \quad \boldsymbol{\Theta}(\mathbf{Z}, \mathbf{U}) \in \mathbb{R}^{p \times m},
\end{align}
where $z_i$ denotes the $i$-th component of the latent state $\mathbf{z}$ evaluated over all time steps, $u_a$ denotes the $a$-th component of the driver vector $\mathbf{u}$, and $p$ is the total number of candidate functions in the library, that is, the number of rows of $\boldsymbol{\Theta}$. All pairwise products $z_i z_j$ and $u_a u_b$ include only distinct combinations with $i \le j$ and $a \le b$, avoiding redundant terms.

The nonlinearity of the identified model is carried entirely by these product terms. The rows $1$, $z_i$, and $u_a$ reproduce an affine model with control; the rows $z_i z_j$, $u_a u_b$, and $z_i u_a$ are what makes the identified dynamics nonlinear, and each has a direct reading in thermospheric terms. The mode--mode products $z_i z_j$ allow the growth rate of one density mode to depend on the amplitude of another, which is how the compressed state carries the difference between how a quiet and an already-disturbed thermosphere respond to heating. The state--driver products $z_i u_a$ let the sensitivity to $F_{10.7}$ and $K_p$ scale with the current state, so the same increment in geomagnetic forcing produces a larger density change at solar maximum than at solar minimum. The driver products $u_a u_b$ carry the joint dependence on solar and geomagnetic forcing that a purely additive model would miss. Because the coefficients $\boldsymbol{\Xi}$ multiplying these fixed functions are what the regression solves for, the identification problem stays linear even though the dynamics it produces are not.

The library is truncated at second order. Second order is the lowest order that admits the state--driver coupling $z_i u_a$ described above, and it keeps the library size at $\mathcal{O}\!\left((r + n_\text{drivers})^2\right)$ terms, which is small enough for the ridge regression to remain well conditioned at the available number of training snapshots. Third-order terms were not tested in this work. They make the library much larger, and the same amount of training data would then have to constrain many more coefficients. The governing dynamics are then cast as the regression problem
\begin{align}
\dot{\mathbf{Z}} = \boldsymbol{\Xi}\,\boldsymbol{\Theta}(\mathbf{Z}, \mathbf{U}),
\end{align}
where $\boldsymbol{\Xi} \in \mathbb{R}^{r \times p}$ is the coefficient matrix to be identified from data. The coefficients are obtained by ridge ($\ell_2$) regression~\cite{marquardt1975ridge},
\begin{align}
\boldsymbol{\Xi} = \dot{\mathbf{Z}}\,\boldsymbol{\Theta}^\top
\!\left(\boldsymbol{\Theta}\,\boldsymbol{\Theta}^\top + \alpha \mathbf{I}\right)^{-1},
\end{align}
where $\alpha > 0$ is a regularization hyperparameter. Ridge regression shrinks the coefficients toward zero without setting any of them exactly to zero, so $\boldsymbol{\Xi}$ is dense in the strict sense and every library term is retained. What the regularization gives is concentration rather than exact sparsity: the identified coefficients cluster on a small subset of the library terms and the rest fall to negligible magnitude. The word sparse is used in that looser sense throughout this paper.

\subsection{Autoregressive Extension (SINDy$_c$-AR)}
\label{subsec:sindy_ar}

The continuous-time formulation above has no memory, which the thermosphere needs because atmospheric heating relaxes over finite timescales and storm-time density enhancements build up and decay gradually. The framework is therefore extended to a discrete-time autoregressive (AR) model, referred to hereafter as SINDy$_c$-AR.

The SINDy$_c$-AR update equation at time step $k$ is
\begin{align}
\mathbf{z}_{k+1} = \mathbf{A}\,\mathbf{z}_k
+ \sum_{j=1}^{n_\text{AR}} \mathbf{A}_j\,\mathbf{z}_{k-j}
+ \mathbf{B}\,\mathbf{u}_k
+ \sum_{j=1}^{n_\text{AR}} \mathbf{B}_j\,\mathbf{u}_{k-j}
+ \boldsymbol{\Xi}\,\boldsymbol{\theta}(\mathbf{z}_k, \mathbf{u}_k)
+ \mathbf{w}_k,
\label{eq:sindy_ar}
\end{align}
where $\mathbf{A} \in \mathbb{R}^{r \times r}$ is the linear state matrix, $\mathbf{A}_j$ captures state-dependent memory at lag $j$, $\mathbf{B}$ and $\mathbf{B}_j$ are the corresponding driver matrices for current and lagged driver values, $\boldsymbol{\Xi}\,\boldsymbol{\theta}(\mathbf{z}_k,\mathbf{u}_k)$ is the nonlinear library term evaluated at the current step, and $\mathbf{w}_k \sim \mathcal{N}(0, \mathbf{Q})$ is process noise. The AR terms $\mathbf{A}_j\mathbf{z}_{k-j}$ and $\mathbf{B}_j\mathbf{u}_{k-j}$ capture explicit temporal memory; lagged state and driver terms are excluded from $\boldsymbol{\theta}$ whenever $n_\text{AR} > 0$ to avoid double-counting memory contributions. The AR order $n_\text{AR}$ is treated as a hyperparameter.

The term $\mathbf{A}\mathbf{z}_k$ is the one-step linear propagator for the current latent state, and it does most of the work in Eq.~\eqref{eq:sindy_ar}. Its eigenvalues set the decay and rotation rates of the retained PCA modes, so on its own it reproduces the relaxation of the density field back toward the training mean once forcing subsides. It is also the term that the DMDc reference model retains, since setting $\mathbf{A}_j = \mathbf{B}_j = \mathbf{0}$ and $\boldsymbol{\Xi} = \mathbf{0}$ in Eq.~\eqref{eq:sindy_ar} recovers $\mathbf{z}_{k+1} = \mathbf{A}\mathbf{z}_k + \mathbf{B}\mathbf{u}_k$ exactly. The remaining terms act as corrections on top of it: $\mathbf{A}_j$ and $\mathbf{B}_j$ add memory beyond one step, and $\boldsymbol{\Xi}\boldsymbol{\theta}$ adds the state- and driver-dependent coupling that a single linear operator cannot represent. In the companion form below, $\mathbf{A}$ occupies the leading block of the first block row of $\mathbf{A}_\text{aug}$ for the same reason.

\subsubsection*{Companion-form state augmentation}
Because the SINDy$_c$-AR model depends on the $n_\text{AR}$ most recent latent states, propagating an error covariance only for $\mathbf{z}_k \in \mathbb{R}^r$ would neglect the cross-correlations between the current and lagged states that arise naturally in an AR process. Following standard state-space practice for autoregressive systems, the filter is implemented on the companion-form lifting of the SINDy$_c$-AR recursion into a Markov state. \citeA{narayanan2025predictive} analyze the predictive capability of Koopman-based DMD algorithms for periodic systems in detail.
In the SINDy$_c$-AR formulation the augmented state vector is defined as
\begin{align}
\boldsymbol{\zeta}_k =
\begin{bmatrix}
\mathbf{z}_k \\ \mathbf{z}_{k-1} \\ \vdots \\ \mathbf{z}_{k-n_\text{AR}}
\end{bmatrix} \in \mathbb{R}^{n_\text{aug}},
\quad n_\text{aug} = r \cdot (n_\text{AR}+1).
\label{eq:aug_state}
\end{align}
In this lifted space the AR recursion~\eqref{eq:sindy_ar} is written as,

\begin{align}
\boldsymbol{\zeta}_{k+1} = \mathbf{F}(\boldsymbol{\zeta}_k,\,\mathbf{u}_k)
+ \mathbf{w}_k^\text{aug},
\label{eq:aug_system}
\end{align}
where $\mathbf{F}(\boldsymbol{\zeta}_k,\mathbf{u}_k)$ denotes the full SINDy$_c$-AR step applied to the augmented state. For covariance propagation the EKF requires the Jacobian of $\mathbf{F}$ with respect to $\boldsymbol{\zeta}_k$, which is derived in Section~\ref{subsec:ekf}. The companion matrix
\begin{align}
\mathbf{A}_\text{aug} =
\begin{bmatrix}
\mathbf{A} & \mathbf{A}_1 & \mathbf{A}_2 & \cdots & \mathbf{A}_{n_\text{AR}} \\
\mathbf{I}_r & \mathbf{0} & \mathbf{0} & \cdots & \mathbf{0} \\
\mathbf{0} & \mathbf{I}_r & \mathbf{0} & \cdots & \mathbf{0} \\
\vdots & & \ddots & & \vdots \\
\mathbf{0} & \cdots & \mathbf{0} & \mathbf{I}_r & \mathbf{0}
\end{bmatrix} \in \mathbb{R}^{n_\text{aug} \times n_\text{aug}}
\label{eq:A_aug}
\end{align}
encodes the current-step dynamics in its top block row and shifts the lag history through identity blocks below the diagonal. The augmented process noise covariance injects uncertainty only into the current-step block,
\begin{align}
\mathbf{Q}_\text{aug} =
\begin{bmatrix}
\mathbf{Q} & \mathbf{0} \\
\mathbf{0} & \mathbf{0}
\end{bmatrix} \in \mathbb{R}^{n_\text{aug} \times n_\text{aug}},
\label{eq:Q_aug}
\end{align}
with $\mathbf{Q} \in \mathbb{R}^{r \times r}$ the process noise covariance of the original latent state. The initial augmented covariance $\mathbf{P}^\text{aug}_{0|0}$ is block-diagonal, with each $r \times r$ block initialized to $\mathbf{P}_{0|0}$.

\subsection{Log-Density Measurement Model}
\label{subsec:meas_model}

Onboard satellite accelerometers and neutral mass spectrometers provide measurements of thermospheric density that span several orders of magnitude. Assimilating them in linear density units would produce poorly scaled filter gains, so they are expressed in $\log_{10}$ density. The latent state $\mathbf{z}_k$ and all dynamics remain in the linear PCA latent space; the log-density formulation applies exclusively to the observation function that maps $\mathbf{z}_k$ to the predicted measurement at the satellite location. Log-density is easier to treat statistically and expresses relative variations directly, which is how density behavior is best characterized~\cite{Emmert2010}.

\subsubsection*{Observation operator}
For a satellite at position $(l, \tau, a)$ (latitude, local time, altitude), let $\boldsymbol{\phi} \in \mathbb{R}^d$ denote the vector of tri-linear interpolation weights at that location. Applying the interpolation to the PCA-reconstructed field gives
\begin{align}
\rho_k = \boldsymbol{\phi}^\top \left(\mathbf{W}\,\mathbf{z}_k + \boldsymbol{\mu}_0\right),
\end{align}
where $\boldsymbol{\mu}_0 \in \mathbb{R}^d$ is the training mean state of Section~\ref{sec:emulation}. Defining the row vector and scalar offset
\begin{align}
\mathbf{H} = \boldsymbol{\phi}^\top \mathbf{W} \in \mathbb{R}^{1 \times r}, \qquad
\mu = \boldsymbol{\phi}^\top \boldsymbol{\mu}_0 \in \mathbb{R},
\label{eq:H_def}
\end{align}
the predicted log-density at the satellite location is
\begin{align}
h_{\log}(\mathbf{z}_k) = \mathbf{H}\,\mathbf{z}_k + \mu,
\label{eq:h_log}
\end{align}
which is affine in the latent state $\mathbf{z}_k$. This affine structure arises from the linear PCA reconstruction and tri-linear spatial interpolation. The measurement equation is
\begin{align}
y_k = \mathbf{H}\,\mathbf{z}_k + \mu + v_k,
\quad v_k \sim \mathcal{N}(0,\,\sigma_v^2),
\label{eq:meas_eq}
\end{align}
where $y_k \in \mathbb{R}$ is the observed $\log_{10}$ density. Assuming Gaussian noise in log space corresponds to multiplicative noise in linear density units, with $\sigma_v^2 \approx (\Delta\rho/\rho)^2/\ln^2(10)$ providing a scale-invariant characterization of measurement uncertainty. The factor $\ln^{-2}(10)$ converts the variance of $\ln\rho$ to that of $\log_{10}\rho$ and is required for consistency with the $\log_{10}$ observation above; the variance depends on the relative error $\Delta\rho/\rho$ alone and not on the absolute density. The log-density formulation implicitly requires that the reconstructed linear density $\boldsymbol{\phi}^\top(\mathbf{W}\mathbf{z}_k + \boldsymbol{\mu}_0)$ remain positive at the interpolated satellite location. We checked this quantity at every filter step along each assimilated and withheld satellite trajectory in the scenarios evaluated; it stayed strictly positive throughout, so no clipping was needed. In broader application of the filter (e.g., extended propagation into unobserved regions at solar minimum or very high altitude), sporadic non-positive reconstructions would need to be flagged and either clipped to a small positive floor or handled by falling back to a linear-space measurement update.

\subsubsection*{Constant Jacobian}
Because $h_{\log}$ is affine in $\mathbf{z}_k$, its Jacobian is the constant matrix
\begin{align}
\mathbf{H}_k = \left.\frac{\partial h_{\log}}{\partial \mathbf{z}}\right|_{\hat{\mathbf{z}}_{k|k-1}} = \mathbf{H},
\label{eq:H_jacobian}
\end{align}
independent of the predicted state. The nonlinear dependence of linear density on $\mathbf{z}_k$ is absorbed into the $\log_{10}$ transform, leaving an affine residual that requires no linearization in the EKF update.

\subsection{EKF Algorithm for Thermospheric Density Estimation}
\label{subsec:ekf}

The EKF operates on the augmented companion state $\boldsymbol{\zeta}_k$ defined in Eq.~\eqref{eq:aug_state}, with the associated augmented covariance $\mathbf{P}^\text{aug}_{k|k} \in \mathbb{R}^{n_\text{aug} \times n_\text{aug}}$. This formulation correctly propagates the cross-covariances between the current and lagged latent states that arise in any AR process. \citeA{narayanan2026stateforecastingestimationframework} describe augmented states in the DMD formulation.
Given a prior estimate $\hat{\boldsymbol{\zeta}}_{k-1|k-1}$ and covariance $\mathbf{P}^\text{aug}_{k-1|k-1}$, the filter proceeds in two steps.

\subsubsection*{Prediction step}
Both the mean state and the covariance are propagated using the full nonlinear basis SINDy$_c$-AR model. The mean prediction follows:
\begin{align}
\hat{\mathbf{z}}_{k|k-1} &= f_\text{SINDy}\!\left(\boldsymbol{\zeta}_{k-1|k-1},\,\mathbf{u}_k\right), \label{eq:ekf_pred_z}\\
\hat{\boldsymbol{\zeta}}_{k|k-1} &=
\begin{bmatrix}
\hat{\mathbf{z}}_{k|k-1} \\
\hat{\mathbf{z}}_{k-1|k-1} \\
\vdots \\
\hat{\mathbf{z}}_{k-n_\text{AR}|k-1}
\end{bmatrix}. \label{eq:aug_pred_z}
\end{align}
The covariance is propagated using the analytical Jacobian $\mathbf{F}_k$ of the full process $\mathbf{F}(\boldsymbol{\zeta},\mathbf{u})$ evaluated at the current state estimate:
\begin{align}
\mathbf{F}_k &= \left.\frac{\partial \mathbf{F}}{\partial \boldsymbol{\zeta}}\right|_{\hat{\boldsymbol{\zeta}}_{k-1|k-1}}
= \mathbf{A}_\text{aug} + \boldsymbol{\Xi}_\text{nl}\,\frac{\partial \boldsymbol{\psi}(\mathbf{z}_k,\mathbf{u}_k)}{\partial \mathbf{z}_k}, \label{eq:F_jac}\\
\mathbf{P}^\text{aug}_{k|k-1} &= \mathbf{F}_k\,\mathbf{P}^\text{aug}_{k-1|k-1}\,\mathbf{F}_k^\top + \mathbf{Q}_\text{aug}. \label{eq:ekf_pred_P}
\end{align}
Here $\boldsymbol{\Xi}_\text{nl} \in \mathbb{R}^{r \times p_\text{nl}}$ is the SINDy coefficient sub-matrix for the nonlinear library block and $\boldsymbol{\psi}(\mathbf{z}_k,\mathbf{u}_k)$ is the standardized form of the library vector $\boldsymbol{\theta}$ entering that block; the derivative $\partial\boldsymbol{\psi}/\partial\mathbf{z}_k$ is computed analytically by differentiating the polynomial feature map of Section~\ref{subsec:sindy_library} with respect to the current latent state. The product $\boldsymbol{\Xi}_\text{nl}\,\partial\boldsymbol{\psi}/\partial\mathbf{z}_k$ is $r \times r$; in Eq.~\eqref{eq:F_jac} it occupies the leading $r \times r$ block of an $n_\text{aug} \times n_\text{aug}$ matrix that is zero elsewhere, in the same pattern as $\mathbf{Q}_\text{aug}$ in Eq.~\eqref{eq:Q_aug}, since only the current-step block of $\boldsymbol{\zeta}_k$ enters the nonlinear library. The Jacobian $\mathbf{F}_k$ therefore captures the first-order Taylor expansion of the SINDy$_c$-AR process at each step, including the contribution of the nonlinear library term, so that the propagation does not rely on inflating the process noise to absorb linearization error. The process noise $\mathbf{Q}_\text{aug}$ is then left to represent stochastic model uncertainty, and Section~\ref{subsec:filter_parameters_tuning} sets it from the model's own one-step prediction residuals and checks it against the observed innovation statistics. If no measurement is available at step $k$, the filter retains $\hat{\boldsymbol{\zeta}}_{k|k-1}$ and propagates $\mathbf{P}^\text{aug}_{k|k-1}$ forward without modification.

\subsubsection*{Update step}
Because measurements observe only the current latent state $\mathbf{z}_k$ (the top $r$ components of $\boldsymbol{\zeta}_k$), the observation operator is augmented as
\begin{align}
\mathbf{H}_\text{aug} = \bigl[\,\mathbf{H} \,\big|\, \mathbf{0}_{1 \times r\,n_\text{AR}}\,\bigr]
\in \mathbb{R}^{1 \times n_\text{aug}},
\label{eq:H_aug_single}
\end{align}
where $\mathbf{H} \in \mathbb{R}^{1 \times r}$ is the log-density observation operator defined in Eq.~\eqref{eq:H_def}. When a $\log_{10}$-density measurement $y_k$ is available, the innovation is
\begin{align}
\nu_k = y_k - \mathbf{H}_\text{aug}\,\hat{\boldsymbol{\zeta}}_{k|k-1} - \mu
      = y_k - \mathbf{H}\,\hat{\mathbf{z}}_{k|k-1} - \mu,
\label{eq:innovation_single}
\end{align}
with physical interpretation
\begin{align}
\nu_k \approx \log_{10}\!\left(\frac{\rho_\text{meas}}{\rho_\text{pred}}\right),
\end{align}
providing a scale-invariant measure of forecast discrepancy. The innovation covariance, Kalman gain, and posterior augmented state are
\begin{align}
\mathbf{S}_k &= \mathbf{H}_\text{aug}\,\mathbf{P}^\text{aug}_{k|k-1}\,\mathbf{H}_\text{aug}^\top + \sigma_{v,k}^2, \\
\mathbf{K}_k &= \mathbf{P}^\text{aug}_{k|k-1}\,\mathbf{H}_\text{aug}^\top\,\mathbf{S}_k^{-1}, \\
\hat{\boldsymbol{\zeta}}_{k|k} &= \hat{\boldsymbol{\zeta}}_{k|k-1} + \mathbf{K}_k\,\nu_k.
\end{align}
The augmented covariance is updated using the Joseph stabilized form to preserve positive semi-definiteness:
\begin{align}
\mathbf{P}^\text{aug}_{k|k} =
\left(\mathbf{I} - \mathbf{K}_k\,\mathbf{H}_\text{aug}\right)
\mathbf{P}^\text{aug}_{k|k-1}
\left(\mathbf{I} - \mathbf{K}_k\,\mathbf{H}_\text{aug}\right)^\top
+ \mathbf{K}_k\,\sigma_{v,k}^2\,\mathbf{K}_k^\top.
\label{eq:joseph}
\end{align}
Because $\mathbf{H}_\text{aug}$ is state-independent (it inherits the constant Jacobian property of $\mathbf{H}$, see Eq.~\eqref{eq:H_jacobian}), the update is exact for the log-density measurement model with no additional linearization required, provided the reconstructed linear density $\boldsymbol{\phi}^\top(\mathbf{W}\mathbf{z}_k + \boldsymbol{\mu}_0)$ remains positive at the interpolated satellite location. In all assimilation scenarios reported here the reconstructed density was verified programmatically to remain strictly positive along every evaluated satellite track; Section~\ref{subsec:meas_model} discusses how to handle the low-density tail in long-duration propagation. The filtered current latent state is extracted from the top block of the augmented posterior,
\begin{align}
\hat{\mathbf{z}}_{k|k} = \hat{\boldsymbol{\zeta}}_{k|k}\bigl[1:r\bigr],
\end{align}
and the associated posterior covariance is the top-left $r \times r$ block $\mathbf{P}_{k|k} = \mathbf{P}^\text{aug}_{k|k}\bigl[1:r,\,1:r\bigr]$.
The filtered latent state is then projected back to the full thermospheric density field via the PCA basis:
\begin{align}
\hat{\mathbf{x}}_{\text{full},k} = \mathbf{W}\,\hat{\mathbf{z}}_{k|k} + \boldsymbol{\mu}_0.
\end{align}

\begin{sidewaysfigure}
\centering
\includegraphics[width=\textwidth]{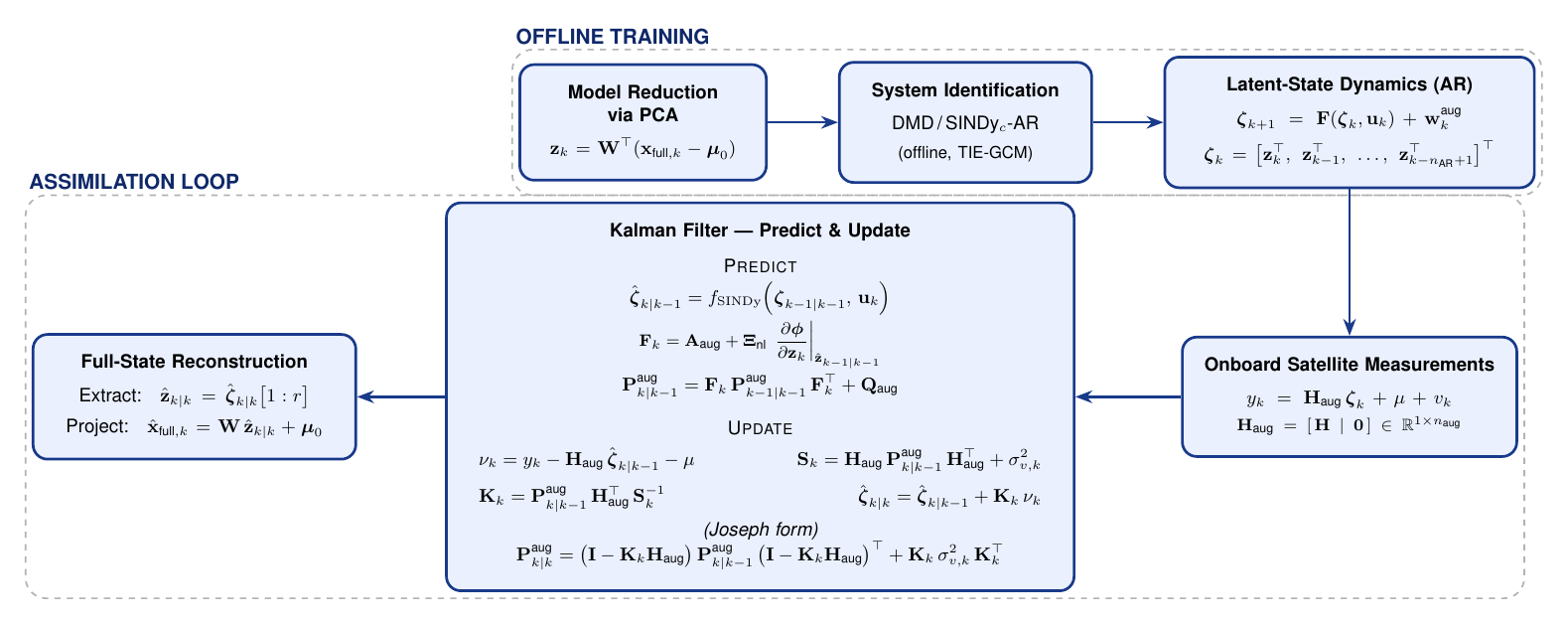}
\caption{Integrated framework for thermospheric density estimation, partitioned into offline training and real-time assimilation phases. 
}
\label{fig:flowchart}
\end{sidewaysfigure}

Figure~\ref{fig:flowchart} illustrates this integrated framework.

\subsection{Multi-Satellite Assimilation in Reduced-Order Space}
\label{subsec:multi_sat_ekf}

Density measurements are often available from several satellites in different orbital planes at once, so the EKF formulation is generalized to $N_s$ satellites.

Each satellite $i$ occupies a distinct orbital position at time step $k$, yielding a per-satellite observation operator and mean offset
\begin{align}
\mathbf{H}^{(i)} = \left(\boldsymbol{\phi}^{(i)}\right)^\top \mathbf{W} \in \mathbb{R}^{1 \times r}, \qquad
\mu^{(i)} = \left(\boldsymbol{\phi}^{(i)}\right)^\top \boldsymbol{\mu}_0 \in \mathbb{R}.
\end{align}
The scalar log-density measurement from satellite $i$ satisfies
\begin{align}
y_k^{(i)} = \mathbf{H}^{(i)}\,\mathbf{z}_k + \mu^{(i)} + v_k^{(i)},
\quad v_k^{(i)} \sim \mathcal{N}\!\left(0,\,\sigma_{v,i}^2\right).
\end{align}
Assuming conditional independence of measurement errors across satellites, all available observations are stacked into an augmented measurement vector
\begin{align}
\mathbf{y}_k^\text{aug} =
\begin{bmatrix}
y_k^{(1)} \\ y_k^{(2)} \\ \vdots \\ y_k^{(N_s)}
\end{bmatrix} \in \mathbb{R}^{N_s},
\end{align}
with diagonal noise covariance
$\mathbf{R}_k^\text{aug} = \mathrm{diag}(\sigma_{v,1}^2,\ldots,\sigma_{v,N_s}^2)$.

The augmented measurement Jacobian is
\begin{align}
\mathbf{H}^\text{aug} =
\begin{bmatrix}
\mathbf{H}^{(1)} \\[4pt] \mathbf{H}^{(2)} \\ \vdots \\ \mathbf{H}^{(N_s)}
\end{bmatrix} \in \mathbb{R}^{N_s \times r},
\end{align}
and the augmented innovation vector is
\begin{align}
\boldsymbol{\nu}_k^\text{aug} = \mathbf{y}_k^\text{aug}
- \mathbf{H}^\text{aug}\,\hat{\mathbf{z}}_{k|k-1}
- \boldsymbol{\mu}^\text{aug},
\end{align}
where $\boldsymbol{\mu}^\text{aug} = [\mu^{(1)},\,\dots,\,\mu^{(N_s)}]^\top$. The augmented innovation vector stacks the $N_s$ scalar innovations defined in Eq.~\eqref{eq:innovation_single}, one per satellite. Its $i$-th entry is the difference between the $\log_{10}$ density measured by satellite $i$ and the $\log_{10}$ density the current forecast predicts at that satellite's own position, so it is the $\log_{10}$ ratio $\log_{10}(\rho^{(i)}_\text{meas}/\rho^{(i)}_\text{pred})$ at that location. Each entry therefore reports, in orders of magnitude, how much the forecast density is off at one point on one orbit, and the filter distributes those $N_s$ mismatches back onto the shared latent state through the Kalman gain. The multi-satellite observation operator is lifted to the companion state space by zero-padding beyond the current-step block,
\begin{align}
\mathbf{H}_{\zeta}^\text{aug} =
\bigl[\,\mathbf{H}^\text{aug} \,\big|\, \mathbf{0}_{N_s \times r\,n_\text{AR}}\,\bigr]
\in \mathbb{R}^{N_s \times n_\text{aug}},
\label{eq:H_aug_multi}
\end{align}
where $\mathbf{H}^\text{aug} \in \mathbb{R}^{N_s \times r}$ is the stacked single-satellite observation matrix. The prediction step uses the analytical-Jacobian covariance propagation~\eqref{eq:ekf_pred_P}; the Kalman gain and posterior update are
\begin{align}
\mathbf{S}_k &= \mathbf{H}_{\zeta}^\text{aug}\,\mathbf{P}^\text{aug}_{k|k-1}\,\left(\mathbf{H}_{\zeta}^\text{aug}\right)^\top + \mathbf{R}_k^\text{aug}, \\
\mathbf{K}_k &= \mathbf{P}^\text{aug}_{k|k-1}\,\left(\mathbf{H}_{\zeta}^\text{aug}\right)^\top \mathbf{S}_k^{-1}, \\
\hat{\boldsymbol{\zeta}}_{k|k} &= \hat{\boldsymbol{\zeta}}_{k|k-1} + \mathbf{K}_k\,\boldsymbol{\nu}_k^\text{aug},
\end{align}
with the Joseph-form covariance update
\begin{align}
\mathbf{P}^\text{aug}_{k|k} =
\left(\mathbf{I} - \mathbf{K}_k\,\mathbf{H}_{\zeta}^\text{aug}\right)
\mathbf{P}^\text{aug}_{k|k-1}
\left(\mathbf{I} - \mathbf{K}_k\,\mathbf{H}_{\zeta}^\text{aug}\right)^\top
+ \mathbf{K}_k\,\mathbf{R}_k^\text{aug}\,\mathbf{K}_k^\top.
\label{eq:joseph_multi}
\end{align}
The updated current latent state $\hat{\mathbf{z}}_{k|k}$ is extracted from the top $r$ components of $\hat{\boldsymbol{\zeta}}_{k|k}$ and projected back to the full density field via $\hat{\mathbf{x}}_{\text{full},k} = \mathbf{W}\hat{\mathbf{z}}_{k|k} + \boldsymbol{\mu}_0$. The computational cost of the EKF update remains tractable: although the augmented state dimension is $n_\text{aug} = r \cdot (n_\text{AR}+1)$, both $r$ and $n_\text{AR}$ are small (10 and 5, respectively), so $n_\text{aug} = 60$ and the dominant cost is the $60 \times 60$ covariance propagation.

\section{Methodology}
\label{sec:methodology}

The satellite datasets, the preprocessing applied to the raw measurements, and the procedure used to tune the EKF parameters are defined below; these choices determine the assimilation results reported in Section~\ref{sec:results}.

\subsection{Data}
\label{subsec:data}

\subsubsection{Reference Model: TIE-GCM}

The reduced-order dynamics model is trained entirely on output from the TIE-GCM \cite{Qian2014}. TIE-GCM output provides the PCA basis $\mathbf{W}$, the mean state $\boldsymbol{\mu}_0$, and the SINDy$_c$-AR coefficient matrices $\mathbf{A}$, $\mathbf{B}$, and $\boldsymbol{\Xi}$. It is not assimilated during filter operation.

\subsubsection{In-Situ Satellite Density Measurements}

Measurements from the following missions are considered: CHAMP, GRACE, GRACE-FO, GOCE, and Swarm A/B/C~\cite{Encarnao2018SwarmAD, visser20161840, vandenijssel20201758, Christian2023, orbits_tudelft}. Each dataset provides density estimates sampled at 10-second intervals along the satellite orbit track; these are decimated onto the 60-second filter grid as described in Section~\ref{subsec:processing_applied}.

Table~\ref{tab:orbit_params} summarizes the orbital parameters of the satellite missions considered in this work. Because all LEO missions experience altitude decay under atmospheric drag, altitudes are reported as the range spanned over each mission's operational period rather than a single scalar.

\begin{table}[t]
\centering
\footnotesize
\setlength{\tabcolsep}{3.5pt}
\caption{Orbital parameters of the satellite missions considered in this work.}
\label{tab:orbit_params}
\begin{tabular*}{\textwidth}{@{\extracolsep{\fill}}lccccc}
\toprule
\textbf{Mission} & \textbf{Operational period} & \textbf{Alt.\ range (km)} & \textbf{Inc.\ (deg)} & \textbf{Ecc.} & \textbf{Period (min)} \\
\midrule
CHAMP     & 2000-07 -- 2010-09 & 456 -- 250 & 87.18 & 0.004  & 93.6 \\
GRACE     & 2002-03 -- 2017-10 & 500 -- 330 & 89.00 & 0.0025 & 94.5 \\
GRACE-FO  & 2018-05 -- present & 490 -- 475 & 89.00 & 0.002  & 94.0 \\
GOCE      & 2009-03 -- 2013-11 & 255 -- 224 & 96.70 & 0.001  & 88.4 \\
Swarm-A   & 2013-11 -- present & 462 -- 430 & 87.35 & 0.001  & 93.7 \\
Swarm-B   & 2013-11 -- present & 511 -- 505 & 87.75 & 0.001  & 94.9 \\
Swarm-C   & 2013-11 -- present & 462 -- 430 & 87.35 & 0.001  & 93.7 \\
\bottomrule
\end{tabular*}
\end{table}

\subsubsection{External Forcing Drivers}

The SINDy$_c$-AR dynamics model requires time series of the solar and geomagnetic forcing indices described in Section~\ref{sec:dataset}. The solar 10.7~cm flux $F_{10.7}$ and its 41-day historical average $\bar{F}_{10.7}$ come from publicly available solar radio records \cite{lisird_data, Yaya17, Dudok14}; the planetary geomagnetic index $K_p$ is the GFZ Potsdam series \cite{Matzka2021}. All three are supplied on the hourly base of the TIE-GCM record, the cadence at which the coefficient matrices were identified, and are interpolated to the 60-second filter time step with a cubic spline, one sub-interval per filter step. The driver vector $\mathbf{u}_k$ includes the current values of $F_{10.7}$, $\bar{F}_{10.7}$, and $K_p$, along with the circular time features $t_1, \ldots, t_4$ defined in Section~\ref{sec:dataset}.

The averaged solar flux driver used here is a 41-day backward-looking mean rather than the 81-day mean centred on the current day that is standard in empirical density models. The centred average requires solar flux values up to 40 days after the epoch being estimated. Those exist in reanalysis but not in the forward-running configuration this filter is intended for, where the centred mean could be supplied only by first forecasting the 40 days of flux that fall after the epoch. Forecast skill for $F_{10.7}$ at that lead time is limited, and the centred driver would therefore carry forecast error into the density estimate, a source of uncertainty the backward-looking driver does not introduce. The 41-day historical mean preserves the slow solar-cycle component that the centred average is meant to supply \cite{tapping2013f107} while remaining causal, so one driver definition serves both a reconstruction and a forecast. It is also the definition the coefficient matrices were identified with: a different averaging window at assimilation time than at identification time would make the driver inconsistent with the fitted coefficients.

\subsection{Preprocessing}
\label{subsec:processing_applied}

Several preprocessing steps are applied to the raw satellite measurements before assimilation.

\subsubsection{Negative density removal}
A few raw accelerometer-derived density estimates are non-physical (negative), typically arising from accelerometer noise during low-drag conditions at high altitude or at solar minimum. These records are removed before any further processing.

\subsubsection{Log-space transformation}
Following the measurement model of Section~\ref{subsec:meas_model}, all density values are converted to $\log_{10}$ units before assimilation. An empirical background density from NRLMSIS~2.1 \cite{Emmert2021, Emmert2022} is computed along the satellite trajectory and stored alongside each measurement for use in benchmark comparisons. For measurements within 2000--2025, a HASDM~\cite{storz2005high} reference density is additionally interpolated along the satellite trajectory (cubic in longitude and latitude, log-PCHIP in altitude, linear in time) and stored alongside each measurement as a higher-fidelity benchmark.

\subsubsection{Temporal alignment}
Each raw measurement is assigned to the filter step that contains it, $k = \lfloor (t - t_0)/T_2 \rfloor$, where $T_2 = 60$~s is the filter step and $t_0$ the filter epoch. Because the density files are sampled at 10~s, six measurements fall within each step; the earliest is retained and the remainder discarded, so that exactly one observation per satellite is assimilated at each filter step. Decimation is applied per satellite \emph{before} the measurement-noise variances of Section~\ref{subsubsec:meas_noise} are computed, so that observations and their variances remain aligned. Retaining all six would treat strongly correlated samples of the same state as independent observations and deflate the effective measurement variance accordingly.

\subsection{Filter Configuration and Parameter Tuning}
\label{subsec:filter_parameters_tuning}

\subsubsection{Time Step and Resolution}

The SINDy$_c$-AR model is identified at an hourly cadence ($T_1 = 3600$~s). The EKF advances on a $T_2 = 60$~s step. The measurements themselves are sampled at 10~s and are decimated to one observation per satellite per step (Section~\ref{subsec:processing_applied}), so the filter step is set by the assimilation cadence rather than by the raw sampling rate. The discrete-time system matrices are transformed from $T_1$ to $T_2$ as
\begin{align}
\mathbf{A}_{T_2} &= \mathbf{A}^{T_2/T_1}, \\
\mathbf{B}_{T_2} &= \left(\mathbf{A}_{T_2} - \mathbf{I}\right)\left(\mathbf{A} - \mathbf{I}\right)^{-1}\mathbf{B},
\end{align}
subdividing each hourly step into $T_1/T_2 = 60$ sub-intervals.

\subsubsection{Reduced-Order Model}
\label{subsubsec:rom}

The reduced-order state is represented by $r = 10$ PCA components. The SINDy$_c$-AR model is configured with $n_\text{AR} = 5$ autoregressive lags. At the hourly identification cadence $T_1$, this spans a 5-hour memory window, which is the timescale over which the thermosphere retains the imprint of a geomagnetic disturbance: storm-time heating raises density within roughly an hour of the forcing onset, and the enhancement then decays over several hours as the disturbance dissipates. Fewer lags truncate that decay and leave the model relying on the current forcing alone; more lags extend the memory past the point where the identified coefficients still carry the storm response, while adding $r$ rows and columns to the augmented state for each additional lag. At $n_\text{AR} = 5$ the augmented dimension is $n_\text{aug} = r(n_\text{AR}+1) = 60$, which keeps the covariance propagation inexpensive. The latent scores themselves point to the same order. Figure~S5 of the Supporting Information shows the autocorrelation and partial autocorrelation of each retained mode over the TIE-GCM record at the hourly cadence. The autocorrelation decays smoothly in every mode, and that is the case in which the partial autocorrelation identifies the order. At lag four three of the ten modes exceed $0.2$ in magnitude, the largest reaching $0.33$. From lag five onward nothing reaches $0.25$: over all ten modes and all lags from five to fifty hours the largest value is $0.21$, and only two of those mode--lag pairs exceed $0.2$ at all. Five lags therefore sit just past the last hour at which the modes carry appreciable direct dependence on their own past. The SINDy library includes polynomial terms up to second order with cross terms $z_i u_a$. Ridge regression is used with $\alpha = 500{,}000$, and all variables are standardized to unit variance before the regression. The regularization strength was selected by a coarse grid search over $\alpha \in \{10^{3}, 10^{4}, 10^{5}, 5\times 10^{5}, 10^{6}\}$ on a held-out TIE-GCM forecast window, choosing the value that minimized open-loop reconstruction error while keeping the identified coefficients small and the regression well conditioned.

\subsubsection{Initial Conditions and Covariance Tuning}
\label{subsubsec:cov_tuning}

Three covariance matrices govern the balance between model forecast and observational correction: the initial state covariance $\mathbf{P}_{0|0}$, the process noise covariance $\mathbf{Q}$, and the measurement noise covariance $\mathbf{R}$. The first two are derived from the identified model rather than chosen by hand. $\mathbf{P}_{0|0}$ is the climatological covariance of the reduced state over the training record, so the filter begins with the spread the modes actually have; its per-mode standard deviations are energy-ordered, from $141.9$ on the leading mode to $3.2$ on the tenth. $\mathbf{Q}$ is the covariance of the one-step prediction residual of the identified SINDy$_c$-AR operator, computed teacher-forced at the hourly identification cadence $T_1$ and scaled to the filter step by $T_2/T_1$. It is a full matrix, not diagonal. Deriving both from the model keeps them on the scale of the state they describe, which a single fixed numerical value cannot do for a basis whose modes span two orders of magnitude in amplitude.

A scalar $q$ multiplying the derived $\mathbf{Q}$ is the only tuned quantity in the filter. It is selected per assimilation window so that the normalized innovation squared, $\text{NIS}_k = \boldsymbol{\nu}_k^\top \mathbf{S}_k^{-1} \boldsymbol{\nu}_k / n_k$ for $n_k$ observations at step $k$, averages close to unity over filtered steps, which is the value a consistent filter returns. Selection uses assimilated orbits only, so the withheld-orbit error remains an independent check rather than the quantity being optimized. The 2003 and 2024 windows are tuned this way, to $q = 1000$ and $q = 600$, giving mean NIS of $1.01$ and $0.98$. The three November 2009 configurations are treated differently. They are compared against one another, and the single-versus-dual generalization result of Section~\ref{subsec:twosat} rests on that comparison, so they share a common $q = 100$ instead of being tuned separately; tuning them individually would confound satellite coverage with process noise. That shared value is not the consistent one for any of them, and their NIS sits above unity. Consistency is traded for comparability here, and the trade is safe for the conclusion drawn: the dual-satellite configuration is the more accurate on the withheld orbit at every shared $q$ from $3$ to $1000$, so the ordering does not depend on where in that range the anchor sits (Supporting Information). All values of $q$ are listed alongside the results in Table~\ref{tab:mape_all}.

$\mathbf{R}$ is not tuned. It stays at the variance implied by the stated 5\% one-sigma sensor error (Section~\ref{subsubsec:meas_noise}), because it describes a measured property of the accelerometers rather than a free parameter. Inflating it is another way to reach $\text{NIS} \approx 1$, and the sweep shows what that buys and what it costs. Scaling $\mathbf{R}$ up by a factor of sixteen in the 2024 scenario reaches consistency at a withheld-orbit MAPE of 16.5\%, against 28.8\% at the unscaled value, but assimilated-orbit MAPE degrades from 2.2\% to 18.6\%: the filter generalizes better because it has largely stopped following the observations it was given. The direction of the withheld-orbit change is not the same in every scenario either: taken at the process noise that restores consistency under the inflated $\mathbf{R}$, the same inflation makes the withheld orbit worse, not better, in both the 2003 and the November 2009 single-satellite cases. We therefore hold $\mathbf{R}$ at its physical value and carry all weighting of model against measurement in $\mathbf{Q}$, and we report this trade rather than selecting on it, since the withheld orbit is the validation quantity. Residual inconsistency between forecast and observation is model error, and model error belongs in $\mathbf{Q}$.

Two properties of this selection matter, because both make an automatic search misleading. First, $\text{NIS} = 1$ is a ridge rather than a point: $\mathbf{R}$ and $\mathbf{Q}$ trade off against one another, and a 210-point sweep over the two found many pairs reaching $\text{NIS} \approx 1$ with assimilated MAPE spanning 0.6 to 18\% within a single scenario, so selecting on $|\log \text{NIS}|$ alone picks an arbitrary point along the ridge. Second, taking the lowest assimilated MAPE inside the consistent band is also degenerate: it decreases monotonically along the ridge to 0.39\%, where a ten-mode model is interpolating the observations rather than representing the field, and NIS cannot detect this because $\mathbf{S}$ grows with $\mathbf{P}$ in step. Holding $\mathbf{R}$ at its physical value and tuning only $q$ removes both degeneracies. A sensitivity analysis over $q$ is given in the Supporting Information.

At the shared November 2009 setting the residual NIS still increases with the number of assimilated satellites, at about 1.9, 3.8 and 6.9 for one, two and three satellites. That increase measures the model's inability to match several orbital regimes at once, and a white-noise $\mathbf{Q}$ cannot properly represent it. The filter is initialized with $\hat{\mathbf{z}}_{0|0} = \mathbf{0}$ and the augmented covariance $\mathbf{P}^\text{aug}_{0|0} \in \mathbb{R}^{n_\text{aug} \times n_\text{aug}}$ block-diagonal, with each $r \times r$ block set to $\mathbf{P}_{0|0}$, reflecting broad initial uncertainty across all lag blocks. The augmented process noise $\mathbf{Q}_\text{aug}$ of Eq.~\eqref{eq:Q_aug} holds the process noise covariance $\mathbf{Q}$, multiplied by the tuned scalar $q$, in its top-left block, with zeros elsewhere. A spin-up period of several hours is used before assimilated results are evaluated. The procedure is deliberately manual rather than innovation-based adaptive; principled online tuning via expectation-maximization or similar methods is deferred to future work (Section~\ref{sec:conclusion}).

\subsubsection{Measurement Noise Covariance}
\label{subsubsec:meas_noise}

The measurement noise variance $\sigma_{v,k}^2$ is estimated using a Monte Carlo approach~\cite{Mehta2018a}. A one-sigma relative sensor error of 5\% in linear density units is assumed, generating $N_\text{MC} = 100$ perturbed samples
\begin{align}
\tilde{\rho}_k^{(s)} = \rho_k^\text{meas}\left(1 + \epsilon^{(s)}\right),
\quad \epsilon^{(s)} \sim \mathcal{N}(0,\,0.05^2).
\end{align}
The variance of $\log_{10}(\tilde{\rho}_k^{(s)})$ across samples gives $\sigma_{v,k}^2$, consistent with $\sigma_v^2 \approx (\Delta\rho/\rho)^2/\ln^2(10)$. Because the same relative error is assumed for every observation, this estimate is in practice constant at $\sigma_v^2 \approx 4.7\times10^{-4}$ in $\log_{10}^2$ units; the Monte Carlo procedure is retained for consistency with \citeA{Mehta2018a} and would yield an observation-dependent variance under a density- or altitude-dependent error model. No additional scaling is applied to $\mathbf{R}$: the value entering the filter is exactly the variance implied by the stated 5\% one-sigma sensor error, and all weighting of model against measurement is carried by $\mathbf{Q}$.

\subsubsection{Experimental Design}
\label{subsec:experimental_design}

Each assimilation experiment is defined by a start and stop date, one or more designated assimilation satellites, and a designated validation satellite whose measurements are withheld from the filter and used only for independent verification. The density files are reported at a 10~s sampling interval while the filter advances on a 60~s step, so one observation per satellite is assimilated at each filter step. Where several satellites are assimilated, their observations at a given step enter as the single stacked update of Section~\ref{subsec:multi_sat_ekf}, with one row of $\mathbf{H}_k^\text{aug}$ per satellite, so that the shared reduced-order state is calibrated by every assimilated orbit simultaneously. The open-loop SINDy$_c$-AR forecast, NRLMSIS~2.1, and HASDM (within its 2000--2025 window) are the baseline comparisons. The open-loop SINDy$_c$-AR forecast is the single no-assimilation reference used throughout; the DMDc model is evaluated under identical conditions as a comparison of assimilated accuracy, and no reduction relative to open-loop is claimed for it. Performance is quantified using MAPE as described in Section~\ref{sec:results}.

\section{Results}
\label{sec:results}

The subsections below report the performance of the SINDy$_c$-AR reduced-order model combined with in situ density measurements through the EKF framework. Five assimilation configurations are evaluated across three geomagnetic periods: two single-satellite cases (CHAMP/GRACE during the 2003 Halloween storm, GRACE-FO/Swarm-C during the May 2024 storm) and three November 2009 configurations of increasing observational coverage (CHAMP only, CHAMP+GRACE, and CHAMP+GRACE+GOCE). The 2003 and 2009 periods fall inside the emulator's 2002--2009 training period; May 2024 lies outside it. The same model structure, reduced-order basis, and filter parameters are used throughout; only the assimilated measurement set varies. Performance is quantified using Mean Absolute Percentage Error (MAPE),
\begin{align}
\mathrm{MAPE} = \frac{100}{N} \sum_{k=1}^{N}
\left|\frac{\hat{\rho}_k - \rho_k^{\mathrm{meas}}}{\rho_k^{\mathrm{meas}}}\right|,
\label{eq:mape}
\end{align}
where $\hat{\rho}_k$ and $\rho_k^{\mathrm{meas}}$ are the estimated and measured densities at time step $k$. MAPE is used because it is scale-invariant across the wide dynamic range of thermospheric density. All five configurations are evaluated against both the open-loop SINDy$_c$-AR forecast and NRLMSIS~2.1 as a physics-based empirical baseline, and additionally against HASDM~\cite{storz2005high} as a higher-fidelity reference for scenarios within its 2000--2025 coverage. A linear DMDc model is also evaluated under the same conditions. Results are best read as showing improvement over the open-loop forecast rather than as a claim of universal superiority over these empirical references, which in some single-satellite cases outperform the assimilated model on withheld orbits when the validation satellite occupies an altitude or local-time regime not well-covered by the assimilated track. Global density maps at 400~km altitude are shown for the two primary scenarios as a qualitative check of large-scale spatial consistency.

\subsection{Single-satellite assimilation: Halloween storm}
\label{subsec:onesat}

The primary single-satellite scenario assimilates CHAMP density measurements during the October--November 2003 Halloween storm (26 Oct -- 7 Nov 2003, $K_p \geq 8$)~\cite{tapley2004gravity, reigber2002champ}, with GRACE withheld for independent validation. This configuration tests generalization across differences in orbital altitude, local time, and satellite characteristics during a geomagnetically extreme interval.

\begin{figure}[!htbp]
\centering
\includegraphics[width=\textwidth]{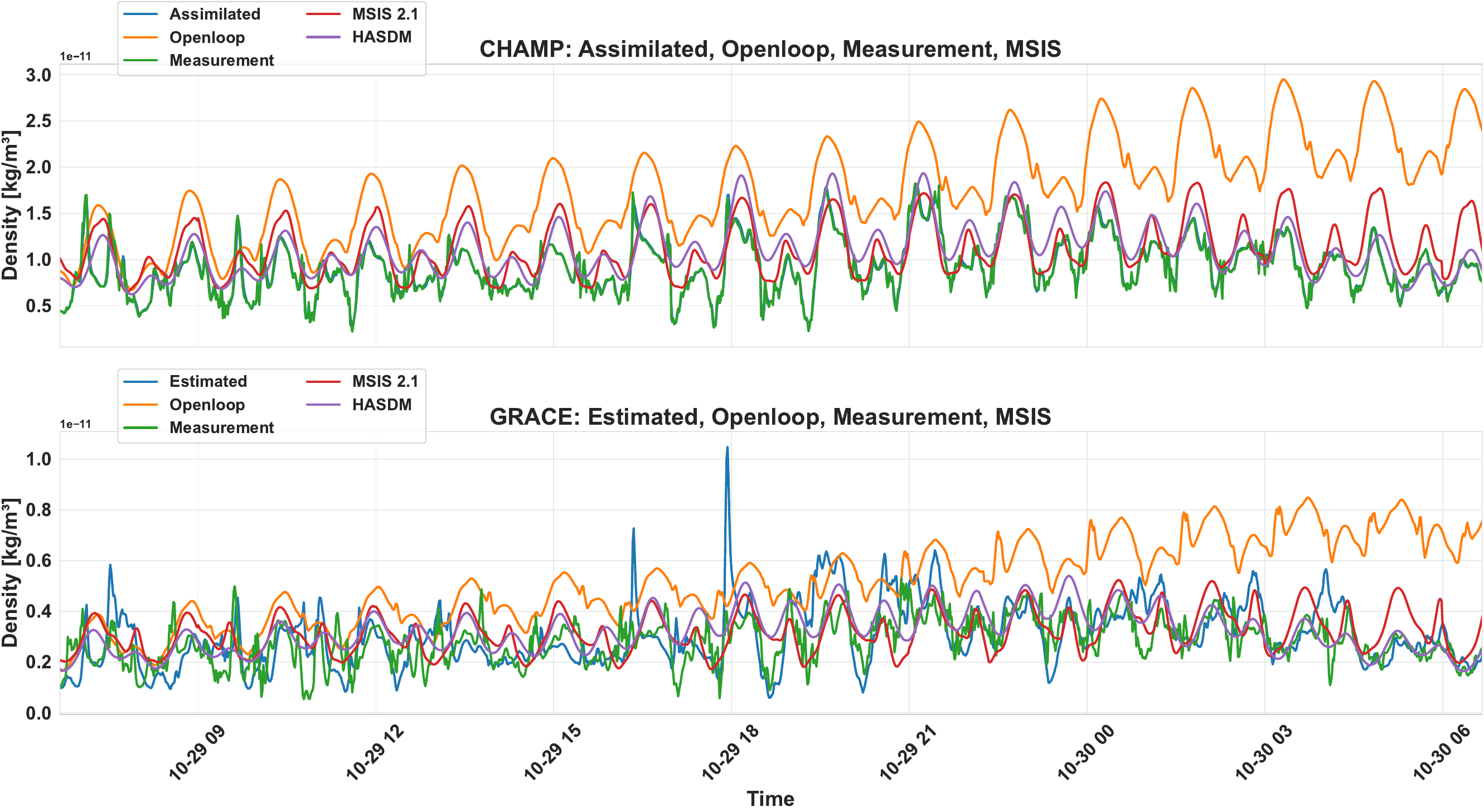}
\caption{Reconstructed thermospheric density during the Halloween storm period compared against independent GRACE measurements. CHAMP observations are assimilated, and GRACE data are withheld for validation. A 24~hour interval, 29--30 October 2003, is shown so that the individual orbital cycles remain resolved; it is the 24~hour window of the assimilation period over which the observed densities are largest. The filter runs over the full 26 October to 7 November window, and the reported errors are computed over that whole period. Short-lived spikes in the lower panel coincide with the largest measurement updates and follow from the shared latent state, not from the withheld track; see Section~\ref{subsec:onesat}.}
\label{fig:sce1_density}
\end{figure}

\begin{figure}[!htbp]
\centering
\includegraphics[width=\textwidth]{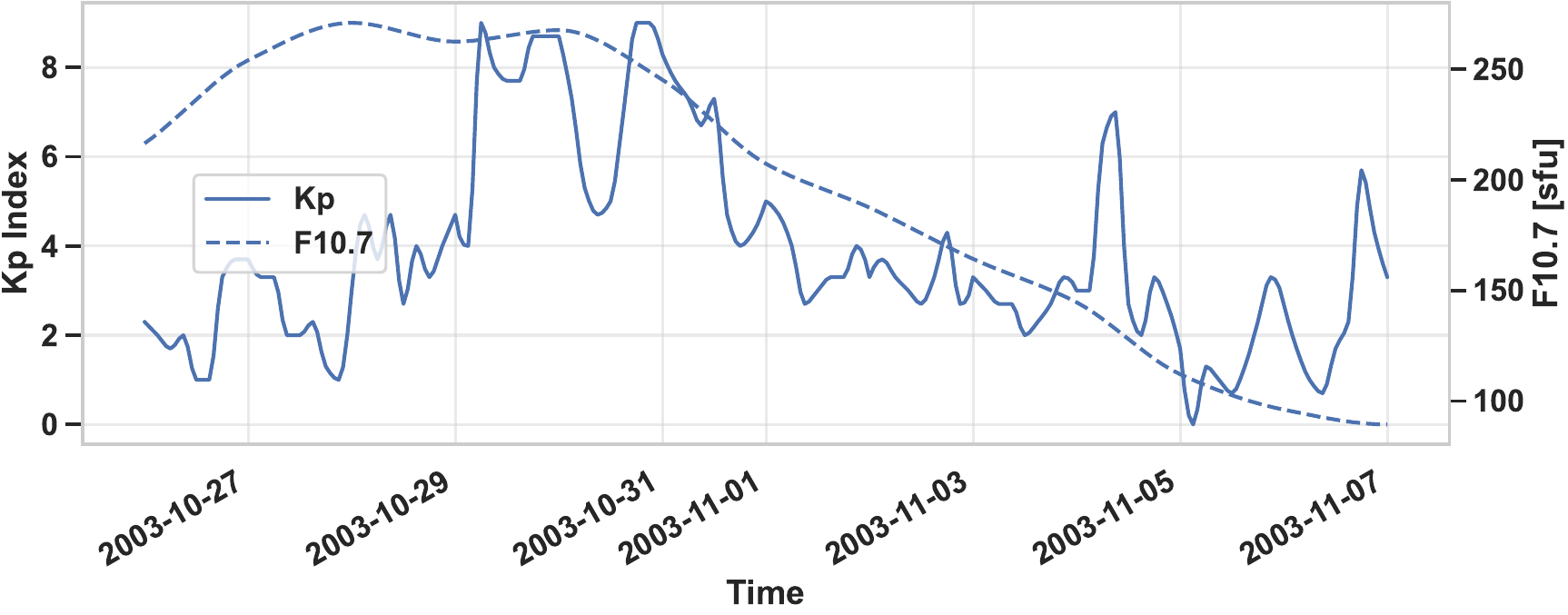}
\caption{Solar and geomagnetic drivers during the Halloween storm.}
\label{fig:sce1_drivers_z_a}
\end{figure}

Figure~\ref{fig:sce1_density} compares the reconstructed density with in situ measurements along both the assimilated CHAMP orbit and the withheld GRACE orbit. For clarity, the results are shown over a 24~hour interval rather than the full assimilation period, so that the individual orbital cycles remain resolved; the interval shown is the 24~hour window over which the observed densities are largest. The filter runs over the full period and all reported errors are computed over it. The period is characterized by sharp, large-amplitude increases in $K_p$ (Figure~\ref{fig:sce1_drivers_z_a}), driving the rapid density enhancements that CHAMP observed during these storms~\cite{Liu2005a}; the same interval produced the largest solar flare on record~\cite{Neil2004}. Despite being calibrated only by CHAMP observations, the assimilated model tracks the timing and magnitude of major storm-time density enhancements with limited phase lag, correcting the open-loop overprediction of peak densities. Validation against GRACE shows a smaller but still large improvement: withheld-orbit MAPE of 22.78\% under $K_p \geq 8$ conditions, compared with 98.86\% for the open-loop forecast, 58.70\% for NRLMSIS~2.1, and 20.04\% for HASDM. Cross-orbit generalization is therefore real but weaker than along the assimilated track, and this is the scenario in which the linear DMDc reference generalizes best (21.46\%, Section~\ref{subsec:mape_summary}). Along the assimilated CHAMP orbit, HASDM gives a MAPE of 20.32\% compared with 48.99\% for NRLMSIS~2.1. The withheld-orbit reconstruction shows short-lived spikes during the storm peak that coincide with the largest measurement updates. At the process noise used here, single-step corrections to the leading mode are large enough to move the reconstructed density on an orbit that was never assimilated, so the spikes are a property of the shared latent state rather than of the GRACE track. They scale with the process noise and disappear at $q = 30$, at a cost of 6.7\% in withheld-orbit MAPE and an NIS of 3.7.

The open-loop error is largest where the measured density peaks. Interpolating the hourly TIE-GCM field along the same tracks, on the same 60~s frame, separates what the parent model contributes from what the reduced-order representation adds. Over the full 2003 window TIE-GCM differs from the measurements by 30.62\% along CHAMP and 28.18\% along GRACE, against 85.92\% and 98.86\% for the open-loop forecast, while the open-loop forecast differs from TIE-GCM by 48.81\% and 68.56\%. Restricted to the 24~hours of Figure~\ref{fig:sce1_density}, the parent model's own error grows to 42.89\% along CHAMP and 34.10\% along GRACE. Part of the peak-time overprediction is therefore inherited. The larger term is the departure of the free-running reduced-order model from TIE-GCM, and since MAPE does not decompose additively, neither figure alone accounts for the 85.92\%. The November 2009 window puts the balance the other way: there the open-loop forecast stays within 10.90\%, 17.06\% and 7.81\% of TIE-GCM along CHAMP, GRACE and GOCE, so almost all of its 59.94\%, 44.95\% and 73.88\% error against the measurements is the parent model's. What separates the two windows is the forcing. Under $K_p \geq 8$ the reduced state leaves the trajectory TIE-GCM follows; at the moderate November 2009 activity it does not. Both sets of values are in Table~\ref{tab:mape_all}. \citeA{sicoli2026sindyar} characterize the emulator against TIE-GCM in more detail, over longer intervals and without an assimilation layer.

Two diagnostics for this scenario are given in the Supporting Information. Figure~S1 shows, for each retained PCA mode, the size of the correction the filter applies relative to the open-loop trajectory of that mode; the corrections concentrate in the first few modes and grow during the storm main phase. Figure~S2 shows the 400~km density field as latitude against local solar time at a single epoch, in four panels: the TIE-GCM field for reference, the open-loop prediction, the assimilated field, and their difference $\rho_\text{assim} - \rho_\text{open}$. The fourth panel is what the assimilation of a single CHAMP track does to the global field, and it shows both the effect of that correction and its limit: the adjustment is largest in the latitude and local-time band the assimilated orbit passes through, and falls off away from it. Both metrics are defined in the Supporting Information captions.

Quantitative results are reported in Table~\ref{tab:mape_all}. Assimilation reduces CHAMP MAPE from 85.92\% (open-loop) to 2.00\%, and GRACE validation MAPE from 98.86\% to 22.78\%, outperforming both the open-loop model and NRLMSIS~2.1. A second single-satellite scenario (GRACE-FO assimilated, Swarm-C withheld, May 2024 storm) tests cross-orbit transferability outside the model training period; results are included in Table~\ref{tab:mape_all} and show consistent improvement, with GRACE-FO MAPE reduced from 94.78\% to 2.19\% and Swarm-C validation MAPE reduced from 94.03\% to 28.83\%. HASDM gives 17.34\% along the assimilated GRACE-FO track and 14.58\% on the withheld Swarm-C orbit, against 2.19\% and 28.83\% respectively for the assimilated reduced-order model. HASDM assimilates a large set of operational drag observations in near-real time and is therefore well constrained through 2024, including along orbits that are withheld from this filter. A single storm interval is not a general characterization of either model.

\subsection{Dual- and multi-satellite assimilation: November 2009}
\label{subsec:twosat}

This scenario targets November 2009, a moderate-activity interval near the declining phase of solar cycle 23 during which CHAMP, GRACE, and GOCE all provide simultaneous in situ density coverage. The concurrent availability of three satellites at distinct altitudes and local-time regimes makes this period a natural testbed for dual- and multi-satellite assimilation and for evaluating cross-orbit generalization.

\begin{figure}[!htbp]
\centering
\includegraphics[width=\textwidth]{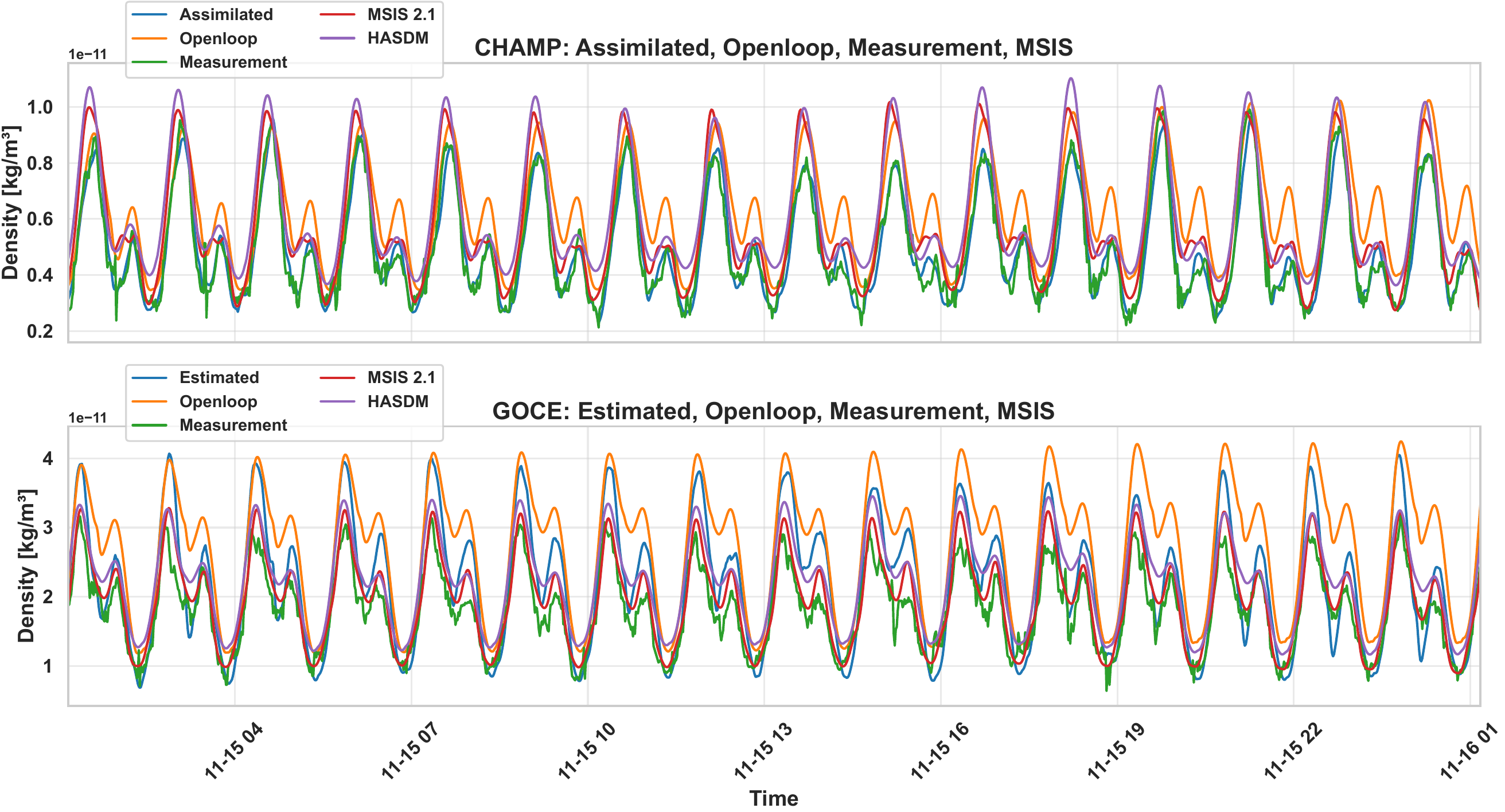}
\caption{Reconstructed thermospheric density during November 2009 compared against GOCE measurements (withheld). CHAMP and GRACE are assimilated simultaneously. A 24~hour interval, 15--16 November 2009, is shown so that the individual orbital cycles remain resolved; it is the 24~hour window of the assimilation period over which the observed densities are largest. The filter runs over the full 14 to 28 November window, and the reported errors are computed over that whole period.}
\label{fig:sce2_dur1_density}
\end{figure}

\begin{figure}[!htbp]
\centering
\includegraphics[width=\textwidth]{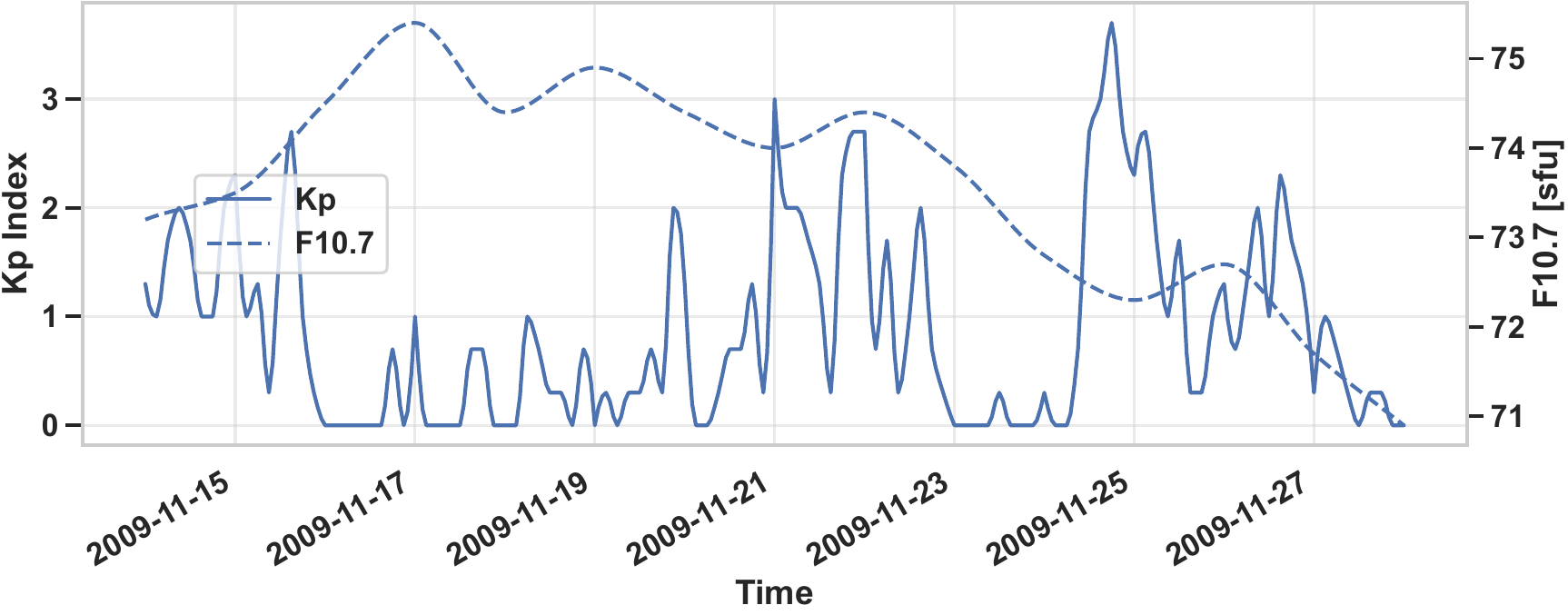}
\caption{Solar and geomagnetic drivers during November 2009.}
\label{fig:sce2_dur1_drivers}
\end{figure}

CHAMP and GRACE are assimilated simultaneously while GOCE is withheld for validation (dual-satellite case). Figure~\ref{fig:sce2_dur1_density} compares the reconstructed densities with measurements along all three satellite tracks; Figure~\ref{fig:sce2_dur1_drivers} shows the November 2009 forcing. The time window is shortened relative to the full assimilation period to improve plot readability. Assimilation reduces residual errors along both assimilated orbits and greatly improves estimates along the withheld GOCE orbit, showing cross-orbit and cross-altitude generalization.

The corresponding diagnostics for this scenario are Figures~S3 and~S4 in the Supporting Information, defined the same way as Figures~S1 and~S2. With two tracks assimilated instead of one, the 400~km difference field of Figure~S4 is adjusted over a wider span of latitude and local time than in the single-satellite case of Figure~S2, which is the spatial counterpart of the MAPE improvement on the withheld GOCE orbit reported below.

As shown in Table~\ref{tab:mape_all}, dual-satellite assimilation reduces CHAMP MAPE from 59.94\% to 6.51\% and GRACE MAPE from 44.95\% to 6.91\%, while GOCE validation MAPE improves from 73.88\% to 23.14\%. Adding GOCE to the assimilation set (multi-satellite, three-satellite case) gives a GOCE MAPE of 13.36\% along its own now-assimilated track. For comparison, a single-satellite baseline (CHAMP only, GOCE withheld) gives a GOCE validation MAPE of 24.15\%. The benefit of the second assimilated orbit for cross-orbit generalization is therefore real but small, about 1\% on the withheld GOCE track, and much smaller than the gain along the assimilated orbits themselves. The margin depends on the process-noise setting, though the ordering largely does not: the dual-satellite configuration is the more accurate at every shared $q$ from $3$ to $1000$, and the ordering reverses only at the two extremes of the range tested, $q = 1$ and $q = 3000$ (Supporting Information).

\subsection{Summary of MAPE across all scenarios}
\label{subsec:mape_summary}

Table~\ref{tab:mape_all} consolidates MAPE results for both the SINDy\textsubscript{c}-AR and DMDc reduced-order models across all assimilation scenarios. Assimilation gives large reductions relative to the open-loop SINDy$_{\mathrm{c}}$-AR forecast in all scenarios. Two conventions apply when reading the table. The open-loop column is the SINDy$_{\mathrm{c}}$-AR forecast and is the single no-assimilation reference used throughout; the DMDc columns report assimilated accuracy under identical filter settings, measurement sets and windows, and are intended for comparison against SINDy$_{\mathrm{c}}$-AR on that basis rather than as reductions relative to this column. The DMDc model propagates on its own free-running trajectory, which is not reported here. 

On assimilated orbits the two models are closely matched. Across all five configurations they differ by no more than 0.33\% on any assimilated orbit. Neither holds a meaningful advantage where the filter is directly constrained by measurements. 

On withheld validation orbits the two also stay close: which model leads varies with the scenario, and no margin exceeds 1.4\%. The per-case values are given in Table~\ref{tab:mape_all}. The two smallest margins, 0.21\% and 0.64\%, are within the run-to-run variability of the Monte Carlo noise estimate. Both models are run with process noise derived and scaled by the same procedure, so the comparison is on equal terms. 

Adding satellites to the assimilation set does not monotonically reduce every individual orbit's MAPE: in the November 2009 multi-satellite case, CHAMP and GRACE MAPEs are higher than in the dual-satellite case, by 1.68\% and 1.09\% respectively, reflecting a redistribution of filter corrections as the shared latent state is calibrated by additional, geometrically distinct measurements. A more direct measure of the same effect is the normalized innovation squared, which rises from about 1.9 to 3.8 to 6.9 as one, two and three satellites are assimilated under an unchanged $\mathbf{R}$ and a shared $q$: the reduced-order model becomes steadily less able to match several orbital regimes at once, even as its accuracy along each individual track remains high. 

The 2024 GRACE-FO to Swarm-C scenario remains the most challenging configuration. At 28.83\% withheld-orbit MAPE it is the largest withheld-orbit error in the study, above the November 2009 values of 24.15\% and 23.14\% and above the 22.78\% of the 2003 storm. The penalty for operating outside the training window is therefore real. In some single-satellite cases, NRLMSIS~2.1 and HASDM outperform the assimilated reduced-order model on withheld orbits, particularly when the validation satellite occupies an altitude or local-time regime not well-covered by the assimilated track. HASDM is included as a higher-fidelity reference for all scenarios; its 2000--2025 record covers every evaluation window, including the 2024 storm.

\begin{sidewaystable}
\centering
\small
\caption{MAPE (\%) summary across all assimilation scenarios for SINDy$_{\mathrm{c}}$-AR and DMDc reduced-order models. ``A'' = assimilated orbit, ``V'' = withheld validation orbit. Open-loop (OL), the parent TIE-GCM field interpolated along the same track, NRLMSIS~2.1 (MSIS) and HASDM values are shown for reference. TIE-GCM output covers 1996--2010, so the 2024 scenario has no TIE-GCM entry. The open-loop column is the SINDy$_{\mathrm{c}}$-AR forecast, which is the single no-assimilation reference used throughout this work; DMDc is included as a comparison of assimilated accuracy only, and its entries are not intended to be read as reductions relative to this column. All November 2009 scenarios share the same open-loop values.}
\label{tab:mape_all}
\setlength{\tabcolsep}{4pt}
\begin{tabular}{llccccccc}
\toprule
\textbf{Scenario} & \textbf{Orbit} & \textbf{Role} & \textbf{SINDy\textsubscript{c}-AR} & \textbf{DMDc} & \textbf{Open-loop} & \textbf{TIE-GCM} & \textbf{MSIS} & \textbf{HASDM} \\
\midrule
\multicolumn{9}{l}{\textit{Single-satellite --- 2003 Halloween storm ($q = 1000$)}} \\
CHAMP to GRACE
  & CHAMP & A &  2.00 &  1.95 & 85.92 & 30.62 & 48.99 & 20.32 \\
  & GRACE & V & 22.78 & 21.46 & 98.86 & 28.18 & 58.70 & 20.04 \\
\midrule
\multicolumn{9}{l}{\textit{Single-satellite --- 2024 geomagnetic storm ($q = 600$)}} \\
GRACE-FO to Swarm-C
  & GRACE-FO & A &  2.19 &  2.09 & 94.78 & --- & 58.47 & 17.34 \\
  & Swarm-C  & V & 28.83 & 29.04 & 94.03 & --- & 57.37 & 14.58 \\
\midrule
\multicolumn{9}{l}{\textit{November 2009 --- single-satellite baseline ($q = 100$)}} \\
CHAMP to GOCE
  & CHAMP & A &  4.68 &  4.49 & 59.94 & 45.78 & 24.54 & 26.14 \\
  & GOCE  & V & 24.15 & 23.51 & 73.88 & 62.86 & 15.66 & 25.83 \\
\midrule
\multicolumn{9}{l}{\textit{November 2009 --- dual-satellite ($q = 100$)}} \\
CHAMP + GRACE to GOCE
  & CHAMP & A &  6.51 &  6.56 & 59.94 & 45.78 & 24.54 & 26.14 \\
  & GRACE & A &  6.91 &  6.76 & 44.95 & 31.02 & 29.19 & 16.54 \\
  & GOCE  & V & 23.14 & 24.29 & 73.88 & 62.86 & 15.66 & 25.83 \\
\midrule
\multicolumn{9}{l}{\textit{November 2009 --- multi-satellite (three satellites) ($q = 100$)}} \\
CHAMP + GRACE + GOCE
  & CHAMP & A &  8.19 &  8.02 & 59.94 & 45.78 & 24.54 & 26.14 \\
  & GRACE & A &  8.00 &  7.81 & 44.95 & 31.02 & 29.19 & 16.54 \\
  & GOCE  & A & 13.36 & 13.69 & 73.88 & 62.86 & 15.66 & 25.83 \\
\bottomrule
\end{tabular}
\end{sidewaystable}

\section{Long-Term Assimilated Thermospheric Density Dataset}
\label{sec:long_term_dataset}

To support reproducibility and enable broader scientific use, a long-term assimilated thermospheric density dataset is released alongside this work. The dataset spans August 2000 through December 2025 and is constructed using all available in situ mass density observations from the candidate LEO satellite missions considered in this work. Satellite availability varies across the 2000--2025 span; the filter ingests whatever subset of missions is active at each epoch.

\begin{figure}[!htbp]
    \centering
    \includegraphics[width=\textwidth]{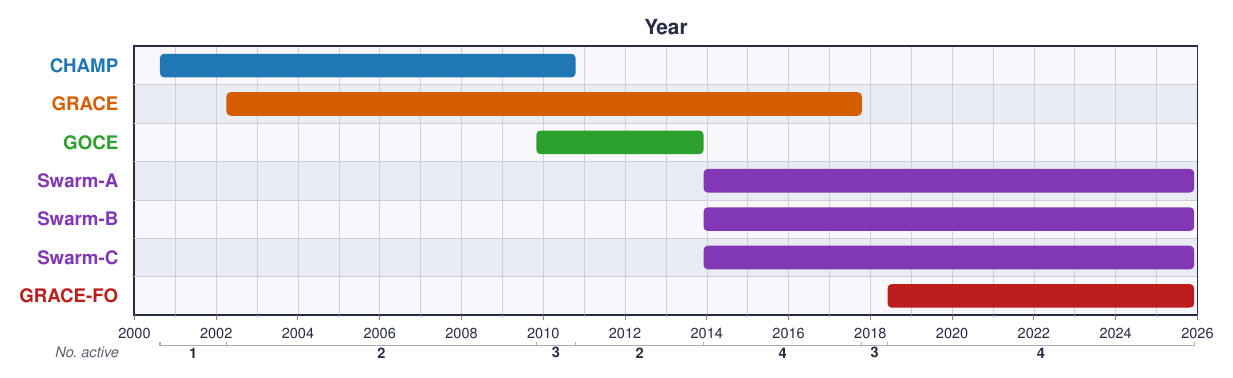}
    \caption{Timeline of satellite missions providing density measurements from August 2000 to December 2025, indicating periods of data availability and overlap for multi-satellite measurement assimilation.}
    \label{fig:sat_assim_timeline}
\end{figure}

The released dataset represents a continuous record composed of single-, dual-, and multi-satellite assimilation periods, reflecting the evolving observational coverage of missions such as CHAMP, GRACE, GRACE-FO, GOCE, and Swarm. When only a single satellite is available, the filter calibrates the reduced-order state using that measurement stream alone. During periods of overlapping missions, multiple satellites are assimilated simultaneously (Figure~\ref{fig:sat_assim_timeline}), improving altitude and local-time coverage of the reconstructed density field. In periods with no available measurements, the dataset defaults to the open-loop reduced-order model propagation.

The released product provides a dynamically consistent, observation-calibrated estimate of thermospheric density that reduces residual bias relative to the open-loop reduced-order model during periods of elevated solar or geomagnetic activity. The multi-satellite assimilation intervals in particular improve cross-orbit and cross-altitude consistency relative to single-satellite calibration. Data are provided in HDF5 format. The record holds the UT epoch of every filter step (\texttt{timestamps}), the posterior reduced-order state (\texttt{z\_filt}), and the free-running open-loop state (\texttt{z\_openloop}), which is the no-assimilation baseline, together with a per-window assimilation summary (\texttt{window\_log}). A per-step diagnostic group (\texttt{steplog}) records which satellites contributed measurements at each step (\texttt{sat\_log}), how many measurements were assimilated (\texttt{n\_meas}), the filter operating mode (\texttt{mode}), and the filter-consistency quantities described below. A companion metadata file records the assimilation configuration: the number of retained components, the filter time step, the start and end dates, the pool of missions ingested, the windowing used for the long-run propagation, and the local-time, latitude, and altitude grid on which the state is reconstructed. The per-step satellite log allows intervals calibrated by several orbits to be separated from intervals with a single orbit or none, which matters when the product is used over periods of changing mission coverage.

The diagnostic group also carries the quantities needed to assess filter behaviour over the full record: the trace of the posterior covariance (\texttt{trace\_P}), its diagonal (\texttt{diag\_P}), the normalized innovation squared (\texttt{NIS}), and the RMS of the innovation vector (\texttt{innov\_rms}) at every step. These are released so that users can judge for themselves where the filter was well constrained and where it was propagating with little observational support.
The dataset is hosted on Zenodo under a CC~BY~4.0 licence~\cite{narayanan2026_thermospheric_dataset}, together with a metadata file documenting the complete variable list, spatial resolution, and temporal cadence, and a reconstruction notebook showing how to recover the density field from the released reduced-order state. The state file is deposited as six sequential parts that concatenate to the original; the record includes a join script and per-part checksums, and the reassembled file is what the notebook expects. Citation information for the dataset is included in the Zenodo record.

\section{Conclusion and Future Work}
\label{sec:conclusion}

\subsection{Summary}

The framework developed here integrates PCA-based dimensionality reduction, autoregressive dynamics identification via SINDy$_c$-AR, and sequential state estimation via an extended Kalman filter. As in the original SINDy formulation~\cite{Brunton2016}, the SINDy$_c$-AR dynamics are linear in the identified coefficients $\boldsymbol{\Xi}$, with nonlinearity entering through the fixed library map $\boldsymbol{\theta}$. Because the AR model depends on multiple lagged states, the filter operates on an augmented companion-form state $\boldsymbol{\zeta}_k = [\mathbf{z}_k^\top, \mathbf{z}_{k-1}^\top, \ldots, \mathbf{z}_{k-n_\text{AR}}^\top]^\top \in \mathbb{R}^{r(n_\text{AR}+1)}$, whose covariance is propagated through the block-companion matrix $\mathbf{A}_\text{aug}$. This augmented propagation correctly captures the cross-lag correlations that a reduced-dimension ($r \times r$) propagation would neglect. The analytical Jacobian of the full SINDy$_c$-AR process, $\mathbf{F}_k = \mathbf{A}_\text{aug} + \boldsymbol{\Xi}_\text{nl}\,\partial\boldsymbol{\psi}/\partial\mathbf{z}_k$, is computed at each step and used to propagate the error covariance consistently with the first-order Taylor expansion, without absorbing the nonlinear library contribution into an inflated process noise. The process noise $\mathbf{Q}$ is therefore intended to represent stochastic model uncertainty; in this work $\mathbf{Q}$ is the covariance of the model's own one-step prediction residual, scaled by a single scalar chosen so that the normalized innovation squared averages close to unity (Section~\ref{subsec:filter_parameters_tuning}). The Joseph stabilized form is used for all covariance updates to guarantee positive semi-definiteness.

The log-density observation model yields an affine measurement function whose Jacobian is state-independent, making the EKF update exact for the measurement step (subject to positivity of the reconstructed linear density at the satellite location, which was satisfied throughout all reported scenarios) and well-conditioned across the full dynamic range of thermospheric density.
The framework was evaluated across three assimilation scenarios during both geomagnetically active and quiet periods. Assimilation consistently reduces density estimation errors along assimilated orbit tracks relative to the open-loop SINDy$_c$-AR forecast. Multi-satellite assimilation improves reconstruction along every assimilated track by calibrating the shared latent state from multiple orbital geometries simultaneously, while its benefit for generalization to withheld orbits is smaller, about 1\% on GOCE between the single- and dual-satellite configurations. Adding GOCE to the November 2009 assimilation set increased CHAMP and GRACE MAPE (from 6.51\% and 6.91\% in the dual-satellite case to 8.19\% and 8.00\% in the three-satellite case), consistent with a redistribution of filter corrections across geometrically distinct measurements (Section~\ref{subsec:twosat}). The nonlinear library formulation and the linear DMDc baseline perform comparably on assimilated and withheld orbits alike (Section~\ref{subsec:mape_summary}). A long-term assimilated density dataset spanning 2000--2025 is released alongside this work.

\subsection{Limitations}

The framework has several limitations. First, the reduced-order model is trained on TIE-GCM output; its accuracy is therefore bounded by the fidelity of TIE-GCM and does not directly incorporate observational climatology. Second, the choice of a fixed $r = 10$ PCA basis is heuristic and may underrepresent mesoscale and high-latitude density structures during extreme storm conditions. Third, $\mathbf{R}$ assumes a uniform 5\% relative measurement error across all missions rather than per-mission uncertainty budgets, which may affect absolute MAPE values. Fourth, while multi-satellite assimilation consistently improves estimates along assimilated tracks, cross-orbit generalization to withheld orbits remains partial. Localized in situ measurements can pull the shared latent state toward a globally consistent correction, but that correction is biased toward the altitude and local-time regions that the assimilated tracks sample; only with adequate spatial and temporal observational coverage does the assimilated state approach a truly self-consistent global solution.
Finally, the physical interpretation of the individual PCA modes, whose filter corrections and residual density fields are shown in Figures~S1--S4 of the Supporting Information, is left to future work. These limitations motivate the future work directions described below.

\subsection{Future Work}

\subsubsection{Extension to higher-fidelity reference environments.}
Scenario-level HASDM comparisons are already reported in Section~\ref{sec:results} as a higher-fidelity empirical reference. The next phase extends the framework to assimilate HASDM-derived density fields directly within the filter and to incorporate TLE-based energy dissipation rate (EDR) density estimates, aligning the inputs with the density products used in operational space situational awareness. TLE-derived densities are attractive because they are available continuously for the full cataloged LEO population across multiple decades, offering the spatial and temporal coverage needed to move from scenario-level assimilation toward a continuously assimilated multi-decade thermospheric density record.

\subsubsection{Richer candidate libraries.}
The library used here is a second-order polynomial expansion in the reduced state and the drivers, chosen for the reasons given in Section~\ref{subsec:sindy_library}. Richer or differently structured libraries, higher reduced ranks and alternative sparsity criteria are all worth testing, in particular for how the choice of library affects accuracy away from the assimilated track.

\subsubsection{Adaptive covariance tuning.}
Although the analytical Jacobian $\mathbf{F}_k$ removes the need for an inflated process noise to compensate for linearization error, the process noise $\mathbf{Q}$ still represents stochastic uncertainty, and the scalar $q$ that scales it is tuned once per window (Section~\ref{subsec:filter_parameters_tuning}). Innovation-based adaptive estimation methods such as expectation-maximization would provide principled online tuning of $\mathbf{Q}$, particularly valuable during storm onset when the stochastic model error structure changes rapidly.

\subsubsection{Ensemble and sigma-point extensions.}
The EKF and an ensemble Kalman filter (EnKF) approximate uncertainty propagation in different ways. The EnKF represents the forecast covariance by a low-rank ensemble estimate, which is subject to sampling error, and obtains it by propagating each member through the nonlinear forecast model. The EKF propagates the covariance explicitly through the model Jacobian. For linear dynamics and a linear observation operator, both reduce to the standard Kalman filter, the EnKF as the ensemble size tends to infinity. For nonlinear dynamics they differ even with an infinite ensemble, because the members keep the nonlinear evolution that a first-order Jacobian drops. Which forecast model the EnKF uses matters. With SINDy$_c$-AR as the forecast model, the 60-dimensional covariance is cheap to propagate explicitly and has no sampling error; that is the computational advantage the EKF uses here, and an EnKF on the same model would trade sampling error for a better treatment of the nonlinear library terms. With a higher-fidelity forecast model such as TIE-GCM, an ensemble is the practical way to propagate uncertainty at all, and the choice is between model fidelity and cost rather than between filters on one model. A UKF would propagate the mean and covariance through the nonlinear library by sigma points rather than by a first-order Taylor expansion, which matters at storm onset when the latent state moves far enough in one step that the linearization is no longer accurate. Separately, a particle or Gaussian-mixture formulation would relax the Gaussian error assumption itself, which is the assumption most in doubt during rapid density enhancements. All of these remain tractable at this state dimension.

\subsubsection{Incorporation of additional measurement types.}
The framework can be extended to other measurement types, such as GPS-based measurements or remote sensing observations of thermospheric composition~\cite{michek_mehta_2026_neutral_species_rope}, provided the measurement can be related to the latent state through a known mapping.

\section*{Open Research Section}

\paragraph*{Data produced by this work.} The assimilated thermospheric density product described in Section~\ref{sec:long_term_dataset}, covering August 2000 to December 2025, is archived on Zenodo under a CC~BY~4.0 licence \cite{narayanan2026_thermospheric_dataset}. The record contains the assimilated and open-loop reduced-order states in HDF5 format, per-step assimilation and filter-consistency diagnostics, a metadata file describing the variables and cadence, and a notebook that reconstructs the full density field from the released state. The state file is 4.9~GB and is deposited as six sequential parts of at most 0.94~GB each, named \texttt{pca\_ark\_results.h5.part-000} through \texttt{part-005}. The parts are a raw byte split rather than an archive format: concatenating them in order reproduces the original file exactly, and the record includes a join script together with SHA-256 checksums for the reassembled file and for each individual part, so that an incomplete transfer can be identified before use.

\paragraph*{Software.} The model code and analysis scripts needed to reproduce every result reported in this manuscript are archived on Zenodo under a CC~BY~4.0 licence \cite{narayanan2026_sindyc_code}.

\paragraph*{Data used as input.} The in situ neutral mass density measurements assimilated and withheld in this work are the CHAMP, GRACE, GRACE-FO, GOCE, and Swarm datasets of \citeA{Christian2023}, obtained from the TU Delft thermosphere database \cite{orbits_tudelft}; the Swarm and GOCE products are described further by \citeA{Encarnao2018SwarmAD}, \citeA{visser20161840}, and \citeA{vandenijssel20201758}. The solar 10.7~cm flux $F_{10.7}$ and its 41-day historical average are obtained from LISIRD \cite{lisird_data} and the composite solar irradiance records of \citeA{Yaya17} and \citeA{Dudok14}. The planetary geomagnetic index $K_p$ is obtained from the GFZ German Research Centre for Geosciences \cite{Matzka2021}. NRLMSIS~2.1 densities used as a benchmark are computed with the models of \citeA{Emmert2021} and \citeA{Emmert2022}. The TIE-GCM thermospheric density fields used to identify the reduced-order model, and used as the reference surface in the global density diagnostics, are available from the NOAA Space Weather Prediction Center on request. The HASDM densities used as a higher-fidelity benchmark are proprietary to Space Environment Technologies and are available directly from SET under their access terms. Neither dataset is redistributable, so neither is included in the archived code and data record; the reproducibility package runs the filter end to end without them, and the code paths that load them are provided but disabled.

\section*{Conflict of Interest Statement}
The authors have no conflicts of interest to disclose.

\acknowledgments

This work was supported in part by the Office of the Director of National Intelligence (ODNI), Intelligence Advanced Research Projects Activity (IARPA), under contract 2023-23060200005. The views and conclusions presented herein are those of the authors and do not necessarily reflect the official policies or endorsements of ODNI, IARPA, or the U.S. Government. This work  is also supported in part by the University of Colorado Boulder (OSC via CU CIRES). The authors also thank the members of the Astrodynamics, Space Science \& Space Technology (ASSIST) lab at WVU for their support.

\bibliography{agusample}

\end{document}